\newif\ifTR
\TRtrue      
 
\documentclass[orivec]{llncs}

\ifTR
\newcommand{\apprefbmsconp}[0]{App.~\ref{sec:BMS-coNP}}
\newcommand{\apprefoptdimbg}[0]{App.~\ref{sec:adfg-bg-complexity}}
\newcommand{\apprefapproxalg}[0]{App.~\ref{sec:approx-alg}}
\else
\newcommand{\apprefbmsconp}[0]{\cite{Ben-AmramG15}}
\newcommand{\apprefoptdimbg}[0]{\cite{Ben-AmramG15}}
\newcommand{\apprefapproxalg}[0]{\cite{Ben-AmramG15}}
\fi

\usepackage{amssymb}
\usepackage{amsmath} 
\usepackage{array,tabularx}
\usepackage{multirow}
\usepackage{paralist}
\usepackage{enumerate}
\usepackage[algoruled,vlined,english]{algorithm2e}
\usepackage[breaklinks]{hyperref}
\usepackage{url}
\usepackage{xspace}
\usepackage[draft]{fixme}
\usepackage{verbatim}
\usepackage{graphicx}
\usepackage{listings}
\spnewtheorem{obs}{Observation}{\bfseries}{\itshape}
\spnewtheorem{prop}{Proposition}{\bfseries}{\itshape}
\spnewtheorem{corol}{Corollary}{\bfseries}{\itshape}

\let\com=\newcommand
\com{\bthm}{\begin{theorem}}
\com{\ethm}{\end{theorem}}
\com{\bdfn}{\begin{definition}}
\com{\edfn}{\end{definition}}
\com{\bobs}{\begin{obs}}
\com{\eobs}{\end{obs}}
\com{\bprop}{\begin{prop}}
\com{\eprop}{\end{prop}}
\com{\blem}{\begin{lemma}}
\com{\elem}{\end{lemma}}
\com{\bcor}{\begin{corol}}
\com{\ecor}{\end{corol}}
\com{\bexm}{\begin{example}}
\com{\eexm}{\end{example}}

\com{\bprf}{\begin{proof}}
\com{\eprf}{\qed\end{proof}}

\com{\bi}{\begin{itemize}}
\com{\ei}{\end{itemize}}
\com{\be}{\begin{enumerate}}
\com{\ee}{\end{enumerate}}

\newcommand{\poly}[1]{{\mathcal #1}}
\newcommand{\intpoly}[1]{{I(#1)}}
\newcommand{\inthull}[1]{{#1}_I}

\newcommand{\convhull}[0]{\mathrm{convhull}}
\newcommand{\cone}[0]{\mathrm{cone}}
\newcommand{\recess}[1]{{\mathcal R_{#1}}}

\newcommand{\trans}[0]{{\mbox{\tiny T}}} 

\newcommand{\transitions}{\poly{Q}}

\newcommand{\trcv}[2]{\ensuremath{\bigl(\begin{smallmatrix}{#1}\hfill\\{#2}\end{smallmatrix}\bigr)}}

\let\tr=\trcv

\newcommand{\rfcoeff}[0]{\lambda}

\let\vect=\vec
\renewcommand{\vec}[1]{\mathbf{#1}}

\newcommand{\tuple}[1]{\langle #1 \rangle}
\newcommand{\set}[1]{\{ #1 \}}

\newcommand{\ints}{\ensuremath{\mathbb Z}\xspace}

\newcommand{\rats}{\ensuremath{\mathbb Q}\xspace}

\let\eps=\varepsilon

\newcommand{\cdont}{{\cdot}}   
\newcommand{\st}[0]{\colon}  

\newcommand{\mlc}[0]{\mbox{MLC}\xspace}

\newcommand{\llrfsym}[0]{\ensuremath{\tau}\xspace}
\newcommand{\diff}[1]{\ensuremath{\Delta #1}}

\newcommand{\lrf}[0]{\mbox{\upshape LRF}\xspace}
\newcommand{\lrfs}[0]{\mbox{{\upshape LRF}s}\xspace}
\newcommand{\llrf}[0]{\mbox{\upshape LLRF}\xspace}
\newcommand{\llrfs}[0]{\mbox{{\upshape LLRF}s}\xspace}
\newcommand{\qlrf}[0]{\mbox{\upshape QLRF}\xspace}
\newcommand{\qlrfs}[0]{\mbox{{\upshape QLRF}s}\xspace}

\newcommand{\llinrfq}[0]{\mbox{\textsc{LexLinRF}\ensuremath{(\rats)}}\xspace}
\newcommand{\llinrfz}[0]{\mbox{\textsc{LexLinRF}\ensuremath{(\ints)}}\xspace}
\newcommand{\llinrfqdim}[0]{\mbox{\textsc{LexLinRF}\ensuremath{(d,\rats)}}\xspace}
\newcommand{\llinrfzdim}[0]{\mbox{\textsc{LexLinRF}\ensuremath{(d,\ints)}}\xspace}

\newcommand{\paramllinrfq}[1]{{#1}{-}\llinrfq}
\newcommand{\paramllinrfz}[1]{{#1}{-}\llinrfz}
\newcommand{\paramllinrfqdim}[1]{{#1}{-}\llinrfqdim}
\newcommand{\paramllinrfzdim}[1]{{#1}{-}\llinrfzdim}
\newcommand{\paramllrf}[1]{{#1}{-}\llrf}
\newcommand{\paramllrfs}[1]{{#1}{-}\llrfs}
\newcommand{\paramqlrf}[1]{{#1}{-}\qlrf}
\newcommand{\paramqlrfs}[1]{{#1}{-}\qlrfs}

\newcommand{\adfg}[0]{\mbox{\upshape ADFG}}
\newcommand{\bg}[0]{\mbox{\upshape BG}}
\newcommand{\bms}[0]{\mbox{\upshape BMS}}

\newcommand{\bmsllinrfq}[0]{\paramllinrfq{\bms}}
\newcommand{\bmsllinrfz}[0]{\paramllinrfz{\bms}}
\newcommand{\bmsllinrfqdim}[0]{\paramllinrfqdim{\bms}}
\newcommand{\bmsllinrfzdim}[0]{\paramllinrfzdim{\bms}}
\newcommand{\bmsllrf}[0]{\paramllrf{\bms}}
\newcommand{\bmsllrfs}[0]{\paramllrfs{\bms}}
\newcommand{\bmsqlrf}[0]{\paramqlrf{\bms}}
\newcommand{\bmsqlrfs}[0]{\paramqlrfs{\bms}}

\newcommand{\bgllinrfqdim}[0]{\paramllinrfqdim{\bg}}
\newcommand{\bgllinrfzdim}[0]{\paramllinrfzdim{\bg}}
\newcommand{\bgllrf}[0]{\paramllrf{\bg}}
\newcommand{\bgllrfs}[0]{\paramllrfs{\bg}}

\newcommand{\adfgllinrfqdim}[0]{\paramllinrfqdim{\adfg}}
\newcommand{\adfgllinrfzdim}[0]{\paramllinrfzdim{\adfg}}
\newcommand{\adfgllrf}[0]{\paramllrf{\adfg}}
\newcommand{\adfgllrfs}[0]{\paramllrfs{\adfg}}

\newcommand{\nollrf}[0]{\mbox{\textsc{None}}\xspace}

\newcommand{\litsum}{\ensuremath{\mathtt{lsum}}}
\newcommand{\sattr}{\ensuremath{\Phi}}
\newcommand{\chtr}[2]{\ensuremath{\Psi}_{{#1}{,}{#2}}}
\newcommand{\anchortr}{{\Omega}}

\newcommand{\redmlcloop}{\ensuremath{{\cal T}}\xspace}
\newcommand{\cmpl}[1]{\ensuremath{\bar{#1}}}

\ifTR
\fi

\begin{document}

\title{Complexity of Bradley-Manna-Sipma Lexicographic Ranking Functions%
\thanks{
This work was funded partially by the EU project FP7-ICT-610582
ENVISAGE: Engineering Virtualized Services
(http://www.envisage-project.eu), by the Spanish MINECO project
TIN2012-38137, and by the CM project S2013/ICE-3006.}
}
\author{Amir M. Ben-Amram\inst{1} \and Samir Genaim\inst{2}
} 
\institute{
School of Computer Science, The Tel-Aviv Academic College, Israel\\
\and
DSIC, Complutense University of Madrid (UCM),  Spain
}

\date{}

\maketitle

\ifTR
\thispagestyle{plain} 
\fi


\begin{abstract}
  In this paper we turn the spotlight on a class of lexicographic
  ranking functions introduced by Bradley, Manna and Sipma in a
  seminal CAV 2005 paper, and establish for the first time the
  complexity of some problems involving the inference of such
  functions for linear-constraint loops (without precondition).  We show that finding such a
  function, if one exists, can be done in polynomial time in a way
  which is sound and complete when the variables range over the
  rationals (or reals). We show that when variables range over the integers, 
  the problem is harder---deciding the existence of a ranking function
  is coNP-complete. 
  Next, we study the problem of minimizing the number of components in
  the ranking function (a.k.a. the dimension). This number is interesting in contexts like computing
  iteration bounds and loop parallelization. Surprisingly, and unlike
  the situation for some other classes of lexicographic ranking
  functions, we find that even deciding whether a two-component
  ranking function exists is harder than the unrestricted problem:
  NP-complete over the rationals and $\Sigma^P_2$-complete over the
  integers.
\end{abstract}


\section{Introduction}
\label{sec:intro}

Proving that a program will not go into an infinite loop is one of the
most fundamental tasks of program verification, and has been the
subject of voluminous research. Perhaps the best known, and often
used, technique for proving termination is the \emph{ranking
  function}.
This is a function $\rho$ that maps the program states into the
elements of a well-founded ordered set, such that $\rho(s) > \rho(s')$
holds for any consecutive states $s$ and $s'$.  This implies
termination since infinite descent in a well-founded order is
impossible.

We focus on \emph{numerical loops}, where a
state is described by the values of a finite set of numerical
variables; we consider the setting of integer-valued variables, as
well as rational-valued (or real-valued) variables.
We ignore details of the programming language; we assume that we
are provided an abstract description of the loop as a finite number of
alternatives, that we call \emph{paths}, each one defined by a finite
set of \emph{linear constraints} on the program variables $x,y,\dots$
and the primed variables $x',y',\dots$ which refer 
to the state
following the iteration.
The following is such a loop consisting of four paths, $\transitions_1,\dots,\transitions_4$:
{\small
\[
\abovedisplayskip=3pt
\belowdisplayskip=3pt
\begin{array}{rllllll}
\transitions_1 =
               &\{ x \ge 0,& x'\le x-1, & & y'=y,      & & z'=z \} \\[-0.3ex]
\transitions_2 =
               &\{ x \ge 0, &x' \le x-1,
               & & y'=y, 
               & z \ge 0,& z' \le z-1 \} \\[-0.3ex]
\transitions_3 = 
               & \{ & x'=x, 
               & y \ge 0,& y'\le y-1,
               & z \ge 0,& z'\le z-1 \} \\[-0.3ex]
\transitions_4 =
               &\{ & x'=x, 
               & y \ge 0,& y'\le y-1,
               & &z'=z \} \\
\end{array}
\]
}
Note that $\transitions_i$ are convex polyhedra. A transition from a
state $\bar{x}$ to $\bar{x}'$ is possible iff $(\bar{x},
\bar{x}')$ is a point in some path $\transitions_i$.
We remark that our results hold for arbitrarily-complex control-flow graphs
(CFGs), we prefer to use the loop setting for clarity.

A popular tool for proving the termination of such loops is
\emph{linear ranking functions} (\lrfs).  An \lrf is a function
$\rho(x_1,\dots,x_n) = a_1x_1+\dots+a_n x_n + a_0$ such that any
transition $(\bar{x},\bar{x}')$ satisfies
\begin{inparaenum}[(i)]
\item\label{intro:lrf1} $\rho(\bar{x}) \ge 0$; and
\item\label{intro:lrf2} $\rho(\bar{x})-\rho(\bar{x}') \ge 1$.
\end{inparaenum}
E.g., $\rho(x,y,z)=x$ is an \lrf for a loop that consists of only
$\transitions_1$ and $\transitions_2$ above, $\rho(x,y,z)=y$ is an
\lrf for $\transitions_3$ and $\transitions_4$, and $\rho(x,y,z)=z$ is
an \lrf for $\transitions_2$ and $\transitions_3$. However, there is
no \lrf that satisfies the above conditions for all paths
$\transitions_1,\ldots,\transitions_4$.
An algorithm to find an \lrf using linear programming (LP) has been
found by multiple researchers in different places and times and in
some alternative
versions~\cite{ADFG:2010,DBLP:conf/tacas/ColonS01,Feautrier92.1,DBLP:journals/tplp/MesnardS08,DBLP:conf/vmcai/PodelskiR04,DBLP:conf/pods/SohnG91}.
Since LP has a polynomial-time complexity, most of these methods yield
polynomial-time algorithms.
These algorithms are complete for loops with rational-valued
variables, but not with integer-valued variables. Indeed, \cite{Ben-AmramG13jv} shows
loops that have \lrfs over the integers but do not even terminate
over the rationals.
In a previous work~\cite{Ben-AmramG13jv} we considered the integer
setting, where complete algorithms were proposed and a complexity
classification was proved: to decide whether an \lrf exists is
\emph{coNP-complete}.

\lrfs do not suffice
for all loops (e.g., the 4-path loop above), and 
thus, a natural question is what to do when an \lrf does not exist;
and a natural answer is to try a richer class of ranking functions. Of
particular importance is the class of
\emph{lexicographic-linear ranking functions} (\llrfs). 
An \llrf
is a $d$-tuple of affine-linear functions,
$\tuple{\rho_1,\dots,\rho_d}$, required to descend lexicographically.
Interestingly, Alan Turing's early demonstration~\cite{Turing48} of
how to verify a program 
used an \llrf for the
termination proof. 
\emph{Algorithms} to find \llrfs for linear-constraint loops (or CFGs)
can use LP techniques, extending the work on \lrfs. 
Alias et al.~\cite{ADFG:2010} extended the polynomial-time \lrf
algorithm to \llrfs and gave a complete solution for  CFGs. As
for \lrfs, the solution is incomplete for integer data, and
in~\cite{Ben-AmramG13jv} we established for \llrfs over the integers
results that parallel those for \lrfs, in particular, to decide
whether an \llrf exists is \emph{coNP-complete}.

Interestingly, when trying to define the requirements from a numeric
``lexicographic ranking function'' (corresponding to the
conditions~\eqref{intro:lrf1} and \eqref{intro:lrf2} on an \lrf,
above), different researchers had come up with different
definitions. In particular, the definition in~\cite{ADFG:2010} is more
restrictive than the definition in~\cite{Ben-AmramG13jv}. Furthermore,
an important paper~\cite{DBLP:conf/cav/BradleyMS05} on \llrf
generation that preceded both works gave yet a different definition.
We give the precise definitions in Sect.~\ref{sec:prelim}; for the
purpose of introduction, let us focus on the 
\llrfs of~\cite{DBLP:conf/cav/BradleyMS05} (henceforth, \bmsllrfs,
after the authors), and illustrate the definition by an example.

Consider the above loop defined by $\transitions_1,\ldots,
\transitions_4$.  A possible \bmsllrf for this loop is 
$\rho(x,y,z) = \tuple{x, y}$.
The justification is this: in $\transitions_1$ and $\transitions_2$,
the function $\rho_1(x,y) = x$ is ranking (non-negative and decreasing
by at least 1). In $\transitions_3$ and $\transitions_4$,
$\rho_2(x,y)=y$ is ranking, while $\rho_1$ is non-increasing.  This is
true over the rationals and \emph{a fortiori} over the integers.
The following points are important: (1) for each path we have an \lrf,
which is one of the components of the \bmsllrf; and (2) previous
(lower-numbered) components are only required to be non-increasing on
that path.  Note that this \llrf does not satisfy the requirements
of~\cite{ADFG:2010} or~\cite{Ben-AmramG13jv}.

The goal of this paper is to understand the \emph{computational
  complexity} of some problems related to \bmsllrfs, starting with the
most basic problem, whether a given loop has such \llrf. We note that
\cite{DBLP:conf/cav/BradleyMS05} does not provide an answer, as a
consequence of attempting to solve a much harder problem---they
consider a loop given with a precondition and search for a \bmsllrf
together with a supporting linear invariant. We do not know if this
problem is even decidable when parameters like the number of
constraints in the invariants are not fixed in advance (when they are,
the approach of~\cite{DBLP:conf/cav/BradleyMS05} is complete, but only
over the reals, and at a high computational cost -- even without a
precondition).

We consider the complexity of finding a \bmsllrf for a given
loop, without preconditions. We prove that this can be done in
polynomial time when the loop is interpreted over the rationals, while
over the integers, deciding the existence of a \bmsllrf is
coNP-complete.  An exponential-time synthesis algorithm is also
given. These results are similar to those obtained for the previously
studied classes of \llrfs~\cite{Ben-AmramG13jv}, but are shown for the
first time for \bmsllrfs.

Next, we consider the number of components $d$ in a \bmsllrf
$\tuple{\rho_1,\dots,\rho_d}$. This number is informally called the
\emph{dimension} of the function.
It is interesting for several reasons: An upper bound on the dimension
is useful for fixing the template in the constraint-solving approach,
and plays a role in analyzing the complexity of corresponding
algorithms.
In addition, an \llrf can be used to infer bounds on the number of
iterations~\cite{ADFG:2010}; assuming linear bounds on individual
variables, a polynomial bound of degree $d$ is clearly implied, which
motivates the desire to minimize the dimension, to obtain tight
bounds. A smaller dimension also means better results when \llrfs are
used to guide parallelization~\cite{Feautrier92.2}.

Importantly, the algorithms of Alias et al.~\cite{ADFG:2010} and
Ben-Amram and Genaim~\cite{Ben-AmramG13jv} 
are optimal w.r.t.\
the dimension, i.e., they 
synthesize \llrfs of minimal dimension for
the respective classes.  We note that it is possible for a loop to have
\llrfs of all three classes but such that the minimal dimension is
different in each (see Sect.~\ref{sec:dimension}).  We also note that,
unlike the case for the previous classes, our synthesis algorithm for
\bmsllrfs is \emph{not} guaranteed to produce a function of minimal
dimension.  This leads us to ask: (1) what is the best \emph{a priori}
bound on the dimension, in terms of the number of variables and paths;
and (2) how difficult it is to find an \llrf of minimal dimension.
As a relaxation of this optimization problem, we can pose the
problem of finding an \llrf that satisfies a given bound on the dimension.
Our results 
are summarized
in Table~\ref{tab:summary}. 
There is a striking difference of
\bmsllrfs from other classes w.r.t.\ to the minimum
dimension problem: the complexity jumps from PTIME (resp. coNP-complete) to NPC (resp. $\Sigma^P_2$-complete) over rationals (resp. integers).
This holds for any fixed dimension larger than one
(dimension one is an \lrf).


\begin{table}[t]
\label{tab:summary}

\centering{
\begin{tabular}{|c|c|c|c|c|c|}
\hline 
\multirow{2}{*}{\llrf type} & \multirow{2}{*}{Dimension bound} & \multicolumn{2}{|c|}{Existence} & \multicolumn{2}{|c|}{Fixed dimension} \\
\cline{3-6} 
                &         & over $\rats$  & over $\ints$ & over $\rats$        & over $\ints$ \\
\hline 
\hline 
ADFG~\cite{ADFG:2010}  
& $\min(n,k)$ &  PTIME & coNP-complete & PTIME & coNP-complete \\ 
\hline \rule{0ex}{3ex}
BG~\cite{Ben-AmramG13jv}      
& $n$         & PTIME  & coNP-complete & PTIME & coNP-complete \\
\hline \rule{0ex}{3ex}
BMS~\cite{DBLP:conf/cav/BradleyMS05} 
& $k$         & PTIME & coNP-complete & NP-complete & $\Sigma^P_2$-complete \\
\hline
\end{tabular}
}

\medskip
\caption{Summary of results, considering a loop of $k$
  paths over $n$ variables. Those in the third row are new, the others are
  from previous works or follow by minor variations.
}

\end{table}


\newcommand{\bsubsec}[1]{
\medskip
\noindent
\textbf{#1}.
}
\section{Preliminaries}
\label{sec:prelim}


\bsubsec{Polyhedra}
A \emph{rational convex polyhedron} $\poly{P} \subseteq \rats^n$
(\emph{polyhedron} for short) is the set of solutions of a set of
inequalities $A\vec{x} \le \vec{b}$, namely $\poly{P}=\{
\vec{x}\in\rats^n \mid A\vec x \le \vec b \}$, where $A \in \rats^{m
  \times n}$ is a rational matrix of $n$ columns and $m$ rows, $\vec
x\in\rats^n$ and $\vec b \in \rats^m$ are column vectors of $n$ and
$m$ rational values respectively.
We say that $\poly{P}$ is specified by $A\vec{x} \le \vec{b}$.
We use calligraphic letters, such as $\poly{P}$ and $\poly{Q}$ to
denote polyhedra.
For a given polyhedron $\poly{P} \subseteq \rats^n$ we let
$\intpoly{\poly{P}}$ be $\poly{P} \cap \ints^n$, i.e., the set of
integer points of $\poly{P}$. The \emph{integer hull} of $\poly{P}$,
commonly denoted by $\inthull{\poly{P}}$, is defined as the convex
hull of $\intpoly{\poly{P}}$.
%
It is known that $\inthull{\poly{P}}$ is also a polyhedron.  An
\emph{integer polyhedron} is a polyhedron $\poly{P}$ such that
$\poly{P} = \inthull{\poly{P}}$. We also say that $\poly{P}$ is
\emph{integral}.

\bsubsec{Multipath Linear-Constraint Loops}
A \emph{multipath} linear-constraint loop (\mlc loop) with $k$
paths has the form:
%
%
$ \bigvee_{i=1}^k A_i\trcv{\vec{x}}{\vec{x}'} \le \vec{c}_i $
%
%
where $\vec{x}=(x_1,\ldots,x_n)^\trans$ and
$\vec{x}'=(x_1',\ldots,x_n')^\trans$ are column vectors, and for 
$q>0$, 
$A_i\in {\rats}^{q\times 2n}$,
$\vec{c}_i\in {\rats}^q$.
%
Each path \(A_i\trcv{\vec{x}}{\vec{x}'} \le \vec{c}_i\) 
is called an \emph{abstract transition}. 
The loop is a \emph{rational loop} if $\vec{x}$ and $\vec{x}'$ range
over $\rats^n$, and it is an \emph{integer loop} if they range over
$\ints^n$.
We say that there is a transition from a state $\vec{x}\in\rats^n$ to
a state $\vec{x}'\in\rats^n$, if for some $1 \le i \le k$, 
\trcv{\vec{x}}{\vec{x}'} 
satisfies the $i$-th abstract transition. 
In such case we say that $\vec{x}$ is an enabled state.
We use $\vec{x}''$ as a shorthand for a transition $\tr{\vec{x}}{\vec{x}'}$,
and consider it as a point in $\rats^{2n}$.
%
The set of transitions satisfying a particular abstract transition 
is a polyhedron in $\rats^{2n}$, denoted $\transitions_i$,  namely
$A_i \vec{x}'' \le \vec{c}_i$. 
%
%
%
In our work it is convenient to represent an \mlc loop by its
transition polyhedra $\transitions_1,\ldots,\transitions_k$, which we often
write with explicit equalities and inequalities.  These are
sometimes referred to as the \emph{paths} of the multipath loop.

\bsubsec{Ranking Functions}
An affine linear function $\rho: \rats^n \mapsto \rats$ is of the form
$\rho(\vec{x}) = \vect{\rfcoeff}\cdot\vec{x} + \rfcoeff_0$ where
$\vect{\rfcoeff}\in\rats^n$ 
and $\rfcoeff_0\in\rats$.
%
We define 
$\diff{\rho}:\rats^{2n}\mapsto\rats$ as
$\diff{\rho}(\vec{x}'')=\rho(\vec{x})-\rho(\vec{x}')$.
Given a set $T\subseteq \rats^{2n}$, representing transitions, we say
that $\rho$ is an \lrf for $T$ if for every $\vec{x}'' \in T$ we have
\eqref{intro:lrf1}  $\rho(\vec{x}) \ge 0$; and \eqref{intro:lrf2} 
$\diff{\rho}(\vec{x}'') \ge 1$.
We say that $\rho$ is an \lrf for a rational (resp. integer) loop, specified by
$\transitions_1,\ldots,\transitions_k$, when it is an \lrf for
$\bigcup_{i=1}^k \transitions_i$ (resp. 
$\bigcup_{i=1}^k
\intpoly{\transitions_i}$).
%
For a rational loop, there is a polynomial-time algorithm to either
find an \lrf or determine that none
exists~\cite{DBLP:conf/vmcai/PodelskiR04}.
 Its essence is that using \emph{Farkas' Lemma}
\cite[p.~93]{Schrijver86}, it is possible to set up an LP problem whose feasibility is equivalent to the existence of
$\rho$ that satisfies \eqref{intro:lrf1}  and \eqref{intro:lrf2}  over $\transitions_1,\ldots,\transitions_k$.

A \emph{$d$-dimensional affine function} $\llrfsym: \rats^n \to
\rats^d$ is expressed by a $d$-tuple $\llrfsym =
\tuple{\rho_1,\dots,\rho_d}$, where each component $\rho_i :\rats^{n}
\to \rats$ is an affine linear function.
The number $d$ is informally called the \emph{dimension} of $\llrfsym$.
Next we define when such a function is
\bmsllrf~\cite{DBLP:conf/cav/BradleyMS05} for a given rational or
integer \mlc loop.  We then compare with 
\adfgllrfs (due to~\cite{ADFG:2010}) and \bgllrfs (due
to~\cite{Ben-AmramG13jv}).

\bdfn[\bmsllrf]
\label{def:bmslexlinearrf}
Given $k$ sets of transitions $T_1, \ldots, T_k \subseteq\rats^{2n}$,
%
we say that $\llrfsym=\tuple{\rho_1,\dots,\rho_d}$ is a \bmsllrf for $T_1, \ldots, T_k$ iff for every $1\le \ell\le k$
there is $1 \le i \le d$ such that the following hold for any
$\vec{x}'' \in T_\ell$:
\abovedisplayskip=1pt
\belowdisplayskip=0pt
\begin{alignat}{ 2 }
 \forall j < i \ .\   && \diff{\rho_j}(\vec{x}'') &\ge 0 \,, \label{eq:bms:llrf1}\\[-0.5ex]
                      && \rho_i(\vec{x}) &\ge 0          \,, \label{eq:bms:llrf2}\\[-0.5ex]
                      && \diff{\rho_i}(\vec{x}'') &\geq 1\,. \label{eq:bms:llrf3} 
\end{alignat}
We say that $T_\ell$ is \emph{ranked by} $\rho_i$.
\edfn

\noindent
We say that $\llrfsym$ is a \bmsllrf for a rational (resp. integer) loop, specified by
$\transitions_1,\ldots,\transitions_k$, when it is a \bmsllrf for
$\transitions_1, \ldots, \transitions_k$ (resp. $\intpoly{\transitions_1}, \cdots, \intpoly{\transitions_k}$).
It is easy to see that the existence of a \bmsllrf implies termination.





\bdfn[\bgllrf]
\label{def:bgllrf}
Given  a set of transitions $T\subseteq\rats^{2n}$,
we say that $\llrfsym=\tuple{\rho_1,\dots,\rho_d}$ is a \bgllrf for
$T$ iff for every $\vec{x}'' \in T$ there is $1 \le i \le d$ such that
the following hold:
\abovedisplayskip=0pt
\belowdisplayskip=0pt
\begin{alignat}{ 2 }
 \forall j < i \ .\   && \diff{\rho_j}(\vec{x}'') &\ge 0 \,, \label{eq:bg:llrf1}\\[-0.5ex]
 \forall j \le i \ .\ && \rho_j(\vec{x}) &\ge 0          \,, \label{eq:bg:llrf2}\\[-0.5ex]
                      && \diff{\rho_i}(\vec{x}'') &\geq 1\,. \label{eq:bg:llrf3} 
\end{alignat}
We say that $\vec x$ is \emph{ranked by} $\rho_i$.
\edfn

\noindent
We say that $\llrfsym$ is a \bgllrf for a rational (resp. integer)
loop, specified by $\transitions_1,\ldots,\transitions_k$, when it is
a \bgllrf for $\transitions_1\cup \cdots\cup \transitions_k$
(resp. $\intpoly{\transitions_1}\cup \cdots\cup
\intpoly{\transitions_k}$).
It is easy to see that the existence of a \bgllrf implies termination.

Note the differences between the definitions: in one sense, \bgllrfs
are more flexible because of the different quantification --- for
every transition $\vec x''$ there has to be a component $\rho_i$ that
ranks it, but $i$ may differ for different $\vec x''$, whereas in
\bmsllrfs, all transitions that belong to a certain $T_\ell$ have to
be ranked by the same component.  In another sense, \bmsllrfs are more
flexible because components $\rho_j$ with $j<i$ can be negative
(compare \eqref{eq:bms:llrf2} with \eqref{eq:bg:llrf2}).  Thus, there
are loops that have a \bmsllrf and do not have a \bgllrf (see loop in
Sect.~\ref{sec:intro}); and vice versa (see~\cite[Ex.~2.12]{Ben-AmramG13jv}).
%
A third type of \llrfs is  attributed to \cite{ADFG:2010}, hence we
refer to it as \adfgllrf. It is similar to \bgllrfs but requires all
components to be non-negative in every enabled state.
That is, condition~\eqref{eq:bg:llrf2} is strengthened.  Interestingly,
the completeness proof in \cite{ADFG:2010} shows that
the above-mentioned flexibility of \bgllrfs adds no power in this
case; therefore, \adfgllrfs are a special case of both \bgllrfs and
\bmsllrfs.

The decision problem \emph{Existence of a \bmsllrf} deals with
deciding whether a given \mlc loop admits a \bmsllrf, we denote it by
\bmsllinrfq and \bmsllinrfz for rational and integer loops
respectively.
The corresponding decision problems for \adfg- and \bgllrfs are solved
in~\cite{ADFG:2010} and~\cite{Ben-AmramG13jv}, respectively, over the
rationals; the case of integers is only addressed
in~\cite{Ben-AmramG13jv} for \bgllrfs, but the complexity results
apply to \adfgllrfs as well.

\section{Synthesis of \bmsllrfs}
\label{sec:synthesis}

In this section we describe a complete algorithm for synthesizing
\bmsllrfs for rational and integer \mlc loops; and show that the
decision problems \bmsllinrfq and \bmsllinrfz are PTIME and
coNP-complete, respectively.
We assume a given \mlc loop $\transitions_1,\ldots,\transitions_k$
where each $\transitions_i$ is given as a set of linear constraints,
over $2n$ variables ($n$ variables and $n$ primed variables).

\bdfn
\label{def:bmsqlrf}
Let $T_1,\ldots,T_k$ be sets of transitions such that
$T_i\subseteq\rats^{2n}$. We say that an affine linear function $\rho$
is a \bms{} quasi-\lrf (\bmsqlrf for short) for $T_1,\ldots,T_k$ if
every transition $\vec{x}''\in T_1 \cup \cdots \cup T_k$ satisfies
$\diff{\rho}(\vec{x}'') \ge 0$, and for at least one $T_\ell$,
$\rho$ is an \lrf (such $T_\ell$ is said to be ranked by $\rho$).
\edfn

\bexm
\label{ex:bms:qlrf}
The following are \bmsqlrfs for the loop consisting of
$\transitions_1,\ldots,\transitions_4$ presented in
Sect.~\ref{sec:intro}:
   $f_1(x,y,z){=}x$, which ranks $\{\transitions_1,\transitions_2\}$;
   $f_2(x,y,z){=}y$ which ranks $\{\transitions_3,\transitions_4\}$; and
   $f_3(x,y,z){=}z$  which ranks $\{\transitions_2,\transitions_3\}$.
\eexm

\blem
\label{lem:rat:bmsqlrfalg}
There is a polynomial-time algorithm that finds a \bmsqlrf $\rho$, if
there is any, for $\transitions_1,\ldots,\transitions_k$.
\elem

\bprf
The algorithm iterates over the paths
$\transitions_1,\ldots,\transitions_k$. In the $i$-th iteration it
checks if there is an \lrf $\rho$ for $\transitions_i$ that is
non-increasing for all other paths, stopping if it finds one. The algorithm makes at most
$k$ iterations. Each iteration can be implemented in polynomial time using
Farkas' Lemma
(as in~\cite{DBLP:conf/vmcai/PodelskiR04}).
\eprf


\begin{algorithm}[t]
\caption{Synthesizing \bmsllrfs}
\label{alg:llrfsyn}
\DontPrintSemicolon
\SetKwFunction{procsyn}{LLRFSYN}
\procsyn{$\tuple{\poly{Q}_1,\ldots,\poly{Q}_k}$}\;
\Begin{
\nl \lIf{$\tuple{\poly{Q}_1,\ldots,\poly{Q}_k}$ are all empty}{\Return \textbf{nil}}\label{alg:terminates}
\nl \If{$\poly{Q}_1,\ldots,\poly{Q}_k$ has a \bmsqlrf $\rho$}{\label{alg:qlrf}
\nl      $\forall 1\le i \le k.~\poly{Q}_i' := \emptyset$ if $\poly{Q}_i$ is ranked by $\rho$, otherwise $\poly{Q}_i'=\poly{Q}_i$\;\label{alg:constraint}
\nl    $\llrfsym \leftarrow \procsyn(\tuple{\poly{Q}'_1,\ldots,\poly{Q}'_k})$\;\label{alg:rec}
\nl    \lIf{$\llrfsym\neq\nollrf$}{\Return  $\rho{::}\tau$}}
\nl  \Return $\nollrf$\label{alg:noqlrf}
}
\end{algorithm}

Our procedure for synthesizing \bmsllrfs is depicted in
Alg.~\ref{alg:llrfsyn}. In each iteration (i.e., call to
\procsyn): it finds a \bmsqlrf $\rho$ for the current 
paths (Line~\ref{alg:qlrf}); it eliminates all paths that are ranked by
$\rho$ (Line~\ref{alg:constraint}); and calls recursively to handle
the remaining paths (Line~\ref{alg:rec}). The algorithm stops when all
paths are ranked (Line~\ref{alg:terminates}), or when it does not find
a \bmsqlrf (Line~\ref{alg:noqlrf}).

\bexm
\label{ex:bms:synth}
Consider the \mlc loop example in Sect.~\ref{sec:intro}.
Procedure \procsyn is first applied to
$\tuple{\transitions_1,\transitions_2,\transitions_3,\transitions_4}$,
and at Line~\ref{alg:qlrf} we can choose the \bmsqlrf $x$
which ranks $\transitions_1$ and $\transitions_2$. Hence these are
eliminated at Line~\ref{alg:constraint}, and at
Line~\ref{alg:rec} \procsyn is applied recursively to
$\tuple{\emptyset,\emptyset,\transitions_3,\transitions_4}$.
Then at Line~\ref{alg:qlrf} we can choose the \bmsqlrf $y$ which ranks
$\transitions_3$ and $\transitions_4$. The next recursive call
receives empty polyhedra, and thus the check at
Line~\ref{alg:terminates} succeeds and the algorithm 
returns~$\tuple{x,y}$.
\eexm

\blem
If $\procsyn(\tuple{\transitions_1,\ldots,\transitions_k})$ returns
$\llrfsym$ different from $\nollrf$, then $\llrfsym$ is a \bmsllrf for
the rational loop $\transitions_1,\ldots,\transitions_k$.
\elem

The proof of the above lemma is straightforward. Thus,
Alg.~\ref{alg:llrfsyn} is a sound algorithm for \bmsllrfs.  The
following proposition shows completeness.

\bprop
\label{prop:rat:llrfcond}
There is a \bmsllrf for $\transitions_1,\ldots, \transitions_k$ if
and only if every subset of $\set{\transitions_1,\ldots,
  \transitions_k}$ has a \bmsqlrf.
\eprop

\bprf
The ``if'' direction is implied by the $\procsyn$ procedure, in such
case it will find a \bmsllrf.
For the ``only if'' direction, let
$\llrfsym=\tuple{\rho_1,\dots,\rho_d}$ be a \bmsllrf for
$\transitions_1,\ldots,\transitions_k$, and let
$\transitions_{\ell_1},\ldots,\transitions_{\ell_j}$ be an arbitrary
subset of the loop's paths.
Since $\llrfsym$ is a \bmsllrf for $\transitions_1,\ldots,
\transitions_k$,  each $\transitions_{\ell_i}$ is ranked by some
$\rho_{l_i}$. Let $l=\min\set{l_1,\ldots,l_j}$, then $\rho_l$ is a
\bmsqlrf for $\transitions_{\ell_1},\ldots,
\transitions_{\ell_j}$.
\eprf

\blem
\label{lem:alg:rat:ptime}
Procedure~$\procsyn$ can be implemented in polynomial time.
\elem

\bprf 
Procedure $\procsyn$ makes at most $k$ steps (since at least one path
is eliminated in every step).
Further, all steps are elementary except checking for 
a \bmsqlrf which
can be done in polynomial time as stated by Lemma~\ref{lem:rat:bmsqlrfalg}.
\eprf

\bcor
\bmsllinrfq $\in$ PTIME. 
\ecor

\noindent
So far we have considered only rational loops, next we consider integer
loops.

\blem
\label{lem:int:bmsqlrfalg}
There is a complete algorithm for synthesizing a \bmsqlrf for
$\intpoly{\transitions_1},\ldots,\intpoly{\transitions_k}$.
\elem

\bprf 
The algorithm  computes the integer hull
$\inthull{\transitions_1}, \ldots, \inthull{\transitions_k}$, 
and then proceeds as in the
rational case (Lemma~\ref{lem:rat:bmsqlrfalg}).
Correctness follows from the fact that for integral polyhedra the 
 implied inequalities over the rationals and integers coincide,
i.e.,  $\inthull{\transitions_1}, \ldots, \inthull{\transitions_k}$ and 
$\intpoly{\transitions_1}, \ldots, \intpoly{\transitions_k}$ have the same \bmsqlrfs.
\eprf

\blem
When procedure \procsyn is applied to the integer hulls
$\inthull{\transitions_1}, \ldots, \inthull{\transitions_k}$, it finds
a \bmsllrf for
$\intpoly{\transitions_1},\ldots,\intpoly{\transitions_k}$, if one
exists.
\elem
\bprf 
Soundness follows from the fact that
$\inthull{\transitions}$ contains $\intpoly{\transitions}$; for completeness, note that:
\begin{inparaenum}[(i)]
\item Prop.~\ref{prop:rat:llrfcond} holds also for integer loops;
and
\item Line~\ref{alg:constraint} of \procsyn does not change the
  transition polyhedra, it only eliminates some, which means that they
  remain integral throughout the recursive calls. Thus, in each
  iteration the check at 
  Line~\ref{alg:qlrf} is complete (see Lemma~\ref{lem:int:bmsqlrfalg}).
\end{inparaenum}
\eprf

In the general case this procedure has an exponential time complexity
since computing the integer hull requires an exponential
time. However, for special cases in which the integer hull can be
computed in polynomial time~\cite[Sect.~4]{Ben-AmramG13jv} it has polynomial
time complexity.  The following lemma implies (assuming P$\ne$NP) that
the exponential time complexity is unavoidable in general.

\bthm 
\label{thm:bms:int:complexity}
\bmsllinrfz is a coNP-complete problem.
\ethm

\bprf 
The coNP-hardness follows from the reduction in~\cite[Sect.~3.1]{Ben-AmramG13jv},
since it constructs a loop that either does not terminate or has an \lrf.
The inclusion in coNP is based on arguments similar to those in
\cite[Sect.~5]{Ben-AmramG13jv}; briefly,
we use the generator representation of the transition polyhedra to
construct a polynomial-size witness against existence of an \llrf
(see~\apprefbmsconp).
\eprf
%


\section{The Dimension of \bmsllrfs} 
\label{sec:dimension}

Ben-Amram and Genaim~\cite[Cor.~5.12, p.~32]{Ben-AmramG13jv} showed
that if a given \mlc loop has a \bgllrf, then it has one of dimension
at most $n$, the dimension of the state
space. The same proof can be used to bound the dimension of \adfgllrfs
by $n$ as well. Hence for \adfgllrfs the bound $\min(n,k)$ holds ($k$
is the number of paths), due to the fact that associating \llrf
components with paths is no loss of generality for
\adfgllrfs~\cite{ADFG:2010}.  In the case of \bmsllrfs, the
bound $k$ clearly holds, and the next example shows that it is tight.

\bexm
\label{ex:bms:maxdim}
Define an \mlc loop $\transitions_1,\ldots,\transitions_k$ for some
$k>0$, over variables $x,y$, where each $\transitions_i=
\{ x' \le x, \ x' + i\cdot y' \le x + i\cdot y - 1, \ x + i\cdot y \ge 0 \}$.
Define $f_i(x,y)=x + i\cdot y$. 
It is easy to check that
\begin{inparaenum}[(i)]
\item\label{ex:bms:maxdim:1} $f_i$ is an \lrf for $\transitions_i$,
and is non-increasing for any $\transitions_j$ with $i<j\le k$; and 
\item\label{ex:bms:maxdim:2}
there are no distinct $\transitions_i$ and $\transitions_j$ that have
a common \lrf.
\end{inparaenum}
From~(\ref{ex:bms:maxdim:1}) it follows that $\tuple{f_1,\ldots,f_k}$
is a \bmsllrf for this loop, and from~(\ref{ex:bms:maxdim:2}) it
follows that any \bmsllrf must have (at least) dimension $k$, since
different paths cannot be ranked by the same component.
We remark that this loop has no \bgllrf (hence, also no \adfgllrf).
\eexm

The above discussion emphasizes the difference between the various
definitions of \llrfs, when considering the dimension. The next example
emphasizes this difference further, it shows that there are loops,
having \llrfs of all three kinds, for which the minimal dimension is
different according to each definition. This also means that the
implied bounds on the number of iterations (assuming, for simplicity,
that all variables have the same upper bound) are different.

\bexm 
\label{ex:differentDim}
Consider an \mlc loop specified by the following paths
\[
\abovedisplayskip=5pt
\belowdisplayskip=5pt
\begin{array}{rllllllll}
\multirow{2}{*}{$\transitions_1 = \Big\{$} 
               & r \ge 0,~~ 
               & 
               & t \ge 0, ~~
               & x \ge 0, ~~
               & 
               & z \ge 0, ~~
               & w \ge 0, 
               & \multirow{2}{*}{$\Big\}$} \\[-0.5ex]
 & r' < r, & & t' < t, & & & & & \\
\multirow{2}{*}{$\transitions_2 = \Big\{$} 
               & r \ge 0, 
               & s \ge 0, ~~
               & t \ge 0, 
               & x \ge 0,
               & 
               & z \ge 0, 
               & w \ge 0, 
               & \multirow{2}{*}{$\Big\}$}\\
 & r' = r, & s' < s, & t' < t, & & & & \\               
\multirow{2}{*}{$\transitions_3 = \Big\{$} 
               & r \ge 0, 
               & s \ge 0, 
               & t' = t
               & x \ge 0, 
               & 
               & z \ge 0, 
               & w \ge 0, 
               & \multirow{2}{*}{$\Big\}$}\\
& r'=r, & s'=s, &  & x' < x, & & & \\
\multirow{2}{*}{$\transitions_4 = \Big\{$} 
               & r \ge 0, 
               & s \ge 0, 
               & t' = t
               & x \ge 0, 
               & y \ge 0, ~~
               & z \ge 0,  
               & w \ge 0, 
               & \multirow{2}{*}{$\Big\}$}\\[-0.5ex]
 & r' = r, & s' = s, & & x' = x, & y' < y, & z' < z, & \\
\multirow{2}{*}{$\transitions_5 = \Big\{$} 
               & r \ge 0, 
               & s \ge 0, 
               & t' = t
               & x \ge 0, 
               & y \ge 0, 
               & z \ge 0,  
               & w \ge 0, 
               & \multirow{2}{*}{$\Big\}$}\\[-0.5ex]
& r' = r, & s' = s, & & x' = x, & y' < y, & z' = z, & w' < w \\
\end{array}
\]
where, for readability, we use $<$ for the relation ``smaller at least
by $1$''.  
This loop has the \bms-\llrf $\tuple{t, x, y}$, which is neither a
\bgllrf or \adfgllrf because $t$ is not lower-bounded on all the
paths.
Its shortest \bgllrf is of dimension $4$, e.g., $\tuple{r, s, x, y}$,
which is not an \adfgllrf because $y$ is not lower-bounded on all the
paths. Its shortest \adfgllrf is of dimension $5$, e.g., $\tuple{r, s, x, z, w}$.
This reasoning is valid for both integer and rational
variables.
\eexm

Next, we consider the problem of minimal dimension. We ask (1) whether our
algorithms return an \llrf with minimal dimension;
and (2) what do we gain (or lose?) in terms of computational tractability if we fix a bound on the dimension in advance.
Importantly,
the algorithms of~\cite{ADFG:2010,Ben-AmramG13jv} 
are optimal w.r.t. the
dimension, i.e., they synthesize \llrfs of minimal dimension. 
In both cases the optimal result is obtained by a greedy algorithm, that constructs the \llrf by adding one
dimension at a time, taking care in
each iteration to rank as many transitions as possible.
The next example shows that a greedy choice in Alg.~\ref{alg:llrfsyn} fails to guarantee optimality,
for both rational and integer loops.
Intuitively, the greedy approach worked in~\cite{ADFG:2010,Ben-AmramG13jv} because the classes of quasi-LRFs used to construact
\llrfs
 are closed under conic combinations, so there is always an
  optimal choice that dominates all others. This
  is not true for \bmsqlrfs. 

\bexm 
Consider the \mlc loop of Sect.~\ref{sec:intro}. If at
Line~\ref{alg:qlrf} Alg.~\ref{alg:llrfsyn}  we seek a \bmsqlrf that ranks a maximal number of
the paths, we can use any of those derived in Ex.~\ref{ex:bms:qlrf}:
$f_1=x$; $f_2=y$; or $f_3=z$.
However, these alternatives lead to
\bms-\llrfs of different dimensions:
\begin{inparaenum}[(i)]
\item choose $f_1$ to rank $\{\transitions_1,\transitions_2\}$, and
  then $f_2$ to rank
  $\{\transitions_3,\transitions_4\}$.
\item choose $f_2$ to rank $\{\transitions_3,\transitions_4\}$, and
  then $f_1$ to rank
  $\{\transitions_1,\transitions_2\}$.
\item choose $f_3$ to rank $\{\transitions_2,\transitions_3\}$, but
  then there is no single function that ranks
  $\{\transitions_1,\transitions_4\}$.
  Take $f_1$ to rank $\transitions_1$ and then $f_2$ to rank
  $\transitions_4$.
\end{inparaenum}
The dimension of the \bms-\llrf in the first two cases  is  $2$, and in
the last one it is  $3$.
\eexm

Since Alg.~\ref{alg:llrfsyn} is not guaranteed to find a \bmsllrf
of minimal dimension, it is natural to ask \emph{how hard is the
  problem of finding a \bmsllrf of minimal dimension?}  This can be
posed as a decision problem: \emph{does a given \mlc loop have a
  \bmsllrf with dimension at most $d$}? This decision problem is
denoted by \bmsllinrfqdim and \bmsllinrfzdim for rational and integer
loops respectively. 
Note that $d$ is a constant, however, it will be clear that accepting $d$
as an input does not change the complexity class of these problems.
Also note that  for $d=1$ it is just the \lrf problem. 
Similar problems can be formulated for \adfg- and \bgllrfs, of
course.  In these two settings, the imposition of a dimension bound
does not change the complexity class.

\bthm
Given a rational \mlc loop, and $d\ge 1$, it is possible to determine
in polynomial time if there is an \adfgllrf (resp.  \bgllrfs) for the
loop of dimension at most $d$. For integer \mlc loops, the problem is
coNP-complete.
\ethm

\bprf
The case of rational loops is straightforward since the corresponding
synthesis algorithms find \llrfs with minimal dimension, and are in PTIME.
The integer case follows easily from the techniques
of~\cite{Ben-AmramG13jv} (see~\apprefoptdimbg).
\eprf


\section{Complexity of \bmsllinrfqdim}
\label{sec:dim:rat}


In this section we show that \bmsllinrfqdim is NP-complete.

\bthm
\label{thm:dim:rat}
For $d \ge 2$, \bmsllinrfqdim is an NP-complete problem.
\ethm

\noindent
For inclusion in NP, a non-deterministic algorithm for the problem
works as follows. First, it \emph{guesses} a partition of
$\{1,\dots,k\}$ into $d$ sets $J_1,\dots,J_d$, of which some may be
empty (we can assume they are last). Then it proceeds
as in Alg.~\ref{alg:llrfsyn} but insists that the paths
indexed by $J_r$ be ranked at the $r$-th iteration. This may fail, and
then the algorithm rejects.  If a \bmsllrf of dimension at most $d$
exists, there will be an accepting computation.

For NP-hardness  we reduce from the NP-complete problem
\emph{d-Colorability of 3-Uniform Hypergraphs}~\cite{Lovasz:hypergraphs1973,PhelpsR84}.
%
An instance of this problem is a set $H$ of $m$ sets $F_1,\ldots,F_m$
(hyperedges, or ``faces''), where each $F_i$ includes exactly $3$
elements from a set of vertices $V = \{1,\dots,n\}$, and we are
asked whether we can choose a color (out of $d$ colors) for each
vertex such that every face is not monocolored.
%

We construct a rational \mlc loop in $3m$ variables and $n$ paths.
%
The variables are indexed by vertices and faces:
variable $x_{i,j}$ corresponds to $i \in F_j \in H$.
For each vertex $1\le i\le n$ we define $\transitions_i$ as a
conjunction of the following:
{\small
\abovedisplayskip=1pt
\belowdisplayskip=2pt
\begin{align} 
 \sum_{k \st i \in F_k} x_{i,k} - \sum_{k \st i \in F_k} x_{i,k}' &\ge 1 & 
\label{eq:iFdec}
\\[-0.1cm]
\sum_{k \st j \in F_k} x_{j,k} - \sum_{k \st j \in F_k} x_{j,k}' &\ge 0  &&
\text{for all vertex $j \ne i$}
\label{eq:iEdec}
\\[-0.1cm]
x_{i,k} &\ge 0 &&
\text{for all face $F_k$ s.t. $i \in F_k$} 
\label{eq:iPos}
\\[-0.1cm]
x_{j,k} &\ge 0 &&
\text{for all vertex $j$ and face $F_k$ s.t. $j\in F_k \land i \notin F_k$} 
\label{eq:jPos}
\\[-0.1cm]
x_{i,k} + x_{j,k} &\ge 0&&
\text{for all vertex $j\ne i$ and face $F_k$ s.t. $i,j\in F_k$}
\label{eq:ijPos}
\end{align}
}
We claim that a rational loop that consists of these $n$ paths has a
\bmsllrf of dimension $d$ iff there is a valid $d$-coloring for the vertices
$V$.

Assume given a $d$-coloring, namely a division of the vertices
in $d$ disjoint sets $V=C_1\cup\cdots\cup C_d$, such that the vertices
of each $C_i$ are assigned the same color.
We construct a \bmsllrf $\tuple{g_1,\ldots,g_d}$ such that
$g_\ell$ ranks all paths $Q_i$ with $i \in C_\ell$. We assume that each
$C_\ell$ is non-empty (otherwise we let $g_\ell(\vec{x})=0$).

We start with $C_1$. For each $F_k\in
H$, define a function $f_k$ as follows: if $F_k \cap C_1 =
\emptyset$ we let $f_k(\vec{x})=0$; if $F_k \cap C_1 = \{i\}$ we let
$f_k(\vec{x}) = x_{i,k}$; and if $F_k \cap C_1 = \{i,j\}$ we let
$f_k(\vec{x}) = x_{i,k}+x_{j,k}$.
We claim that $g_1(\vec{x}) = \sum_{k} f_k$ is a \bmsqlrf for
$\transitions_1,\ldots,\transitions_n$ that ranks all paths
$\transitions_i$ with $i \in C_1$, which we justify as follows:
\smallskip
\begin{compactenum}

\item $g_1$ is non-increasing on all $\transitions_j$, and decreasing
  for each $\transitions_{i}$ with $i\in C_1$. To see this, rewrite
  $g(\vec{x})$ as $\sum_{\iota\in C_1} \sum_{k \st \iota\in F_k}
  x_{\iota,k}$.  As each inner sum is non-increasing
  by~(\ref{eq:iFdec},\ref{eq:iEdec}), we conclude that $g_1$ is
  non-increasing on all paths.  Moreover, for $i\in C_1$, the sum
  $\sum_{k \st i\in F_k} x_{i,k}$ appears in $g_1$ and is decreasing
  according to \eqref{eq:iFdec}, thus $g_1$ is decreasing for each
  $\transitions_{i}$ with $i\in C_1$.

\item $g_1$ is non-negative for all $\transitions_i$ with $i \in C_1$,
  because all $f_k$ are non-negative on these paths. 
To see this, pick
  an arbitrary $i \in C_1$ and an arbitrary face $F_k$:
  if $i \in F_k$, and it is the only vertex from $C_1$ in $F_k$, then
  $f_k(\vec{x})=x_{i,k}$ is non-negative on $\transitions_i$
  by~\eqref{eq:iPos};
  if $i \in F_k$ but there is another vertex $j\in C_1$ in $F_k$, then
  $f_k(\vec{x}) = x_{i,k} + x_{j,k}$ is non-negative on
  $\transitions_i$ by~\eqref{eq:ijPos};
  if $i \notin F_k$, then for any $j \in F_k$ we have $x_{j,k} \ge 0$
  by~\eqref{eq:jPos}, and then $f_k$ is non-negative since it is a sum
  of such variables. 
  Note that $g_1$ can be negative for $\transitions_j$ with $j\not\in
  C_1$.
\end{compactenum}
Similarly, we construct \bmsqlrfs $g_2,\ldots,g_d$ such that $g_\ell$
ranks $\transitions_i$ for $i\in C_\ell$. Clearly
$\tuple{g_1,\ldots,g_d}$ is a \bmsllrf for this loop.

Now suppose we have a \bmsllrf of dimension $d$; we analyze what paths
$\transitions_i$ can be associated with each component, and show that
for any face $F_k$, the three paths that are indexed by its vertices,
i.e., $\transitions_i$ for $i\in F_k$, cannot be all associated with
the same component. Which clearly yields a $d$-coloring.

Suppose that for some face $F_k = \{i_1,i_2,i_3\}$, the paths
$\transitions_{i_1}, \transitions_{i_2}$ and $\transitions_{i_3}$ are
associated with the same component, i.e., all ranked by the same
function, say $g$.
Thus $\diff g(\vec{x}'') \ge 1$ must be implied by the
constraints of $\transitions_{i_1}, \transitions_{i_2}$ and
$\transitions_{i_3}$, independently.
Now since, in each path, the only constraint with a non-zero free
coefficient is~\eqref{eq:iFdec}, it follows that the coefficients of
variables $x_{i_1,k}$, $x_{i_2,k}$ and $x_{i_3,k}$ in $g(\vec{x})$ are
positive, i.e., $g(\vec{x}) = a_1\cdot x_{i_1,k} + a_2\cdot x_{i_2,k}
+ a_3\cdot x_{i_3,k} + h(\vec{x})$ where $h(\vec{x})$ is a combination
of other variables, and $a_1,a_2,a_3 > 0$.
Similarly, $g(\vec{x}) \ge 0$ must be implied by the
constraints of each of three paths independently.
For this to hold, $g$ must be a positive linear combination of
functions constrained to be non-negative by these paths, and do not
involve primed variables.
Now consider variables $x_{i_1,k}$, $x_{i_2,k}$ and $x_{i_3,k}$, and
note that they participate only in the following constraints in
$\transitions_{i_1}$ (left), $\transitions_{i_2}$ (middle) and
$\transitions_{i_3}$ (right):
{\small
\[
\abovedisplayskip=1pt
\belowdisplayskip=1pt
\begin{array}{rlcrlcrl}
x_{i_1,k} &\ge 0  &~~~~~~~~~&
x_{i_2,k} &\ge 0  &~~~~~~~~~&
x_{i_3,k} &\ge 0  \\[-0.5ex]

x_{i_1,k} + x_{i_2,k} &\ge 0 &&
x_{i_1,k} + x_{i_2,k} &\ge 0 &&
x_{i_2,k} + x_{i_3,k} &\ge 0 \\[-0.5ex]

x_{i_1,k} + x_{i_3,k} &\ge 0 &&
x_{i_2,k} + x_{i_3,k} &\ge 0 &&
x_{i_1,k} + x_{i_3,k} &\ge 0 
\end{array}
\]
}
This means that the corresponding coefficients in $g$, i.e.,
$\bar{a}=(a_1~a_2~a_3)$, must be equal to linear combinations of the
corresponding coefficients in the above constraints. Namely, there
exist $b_1,\ldots, b_9 \ge 0$ such that
{\small
\[
\abovedisplayskip=1pt
\belowdisplayskip=1pt
\begin{array}{ccccc}
\bar{a}
 =
\begin{pmatrix}
b_1 & b_2 & b_3 
\end{pmatrix}
\cdot
\begin{pmatrix}
1 & 0 & 0 \\[-0.6ex]
1 & 1 & 0 \\[-0.6ex]
1 & 0 & 1
\end{pmatrix}
&~~~&
\bar{a}
 =
\begin{pmatrix}
b_4 & b_5 & b_6 
\end{pmatrix}
\cdot
\begin{pmatrix}
0 & 1 & 0 \\[-0.6ex]
1 & 1 & 0 \\[-0.6ex]
0 & 1 & 1
\end{pmatrix}
&~~~&
\bar{a}
 =
\begin{pmatrix}
b_7 & b_8 & b_9 
\end{pmatrix}
\cdot
\begin{pmatrix}
0 & 0 & 1 \\[-0.6ex]
0 & 1 & 1 \\[-0.6ex]
1 & 0 & 1
\end{pmatrix}
\end{array}
\]
}
From these nine equations, and the constraints $b_i \ge 0$ for all
$i$, we necessarily get $a_1 = a_2 = a_3 = 0$, which contradicts
$a_1,a_2,a_3>0$ as we concluded before, and thus paths corresponding
to $\{i_1,i_2,i_3\}$ of $F_k$ cannot be all associated with the same
component. This concludes the proof of Th.~\ref{thm:dim:rat}.
%
%




\newcommand{\bintcs}[0]{\ensuremath{\cal F}}
\newcommand{\EAsentence}{($\star$)\xspace}

\section{Complexity of \bmsllinrfzdim}
\label{sec:dim:int}

In this section we turn to the problem \bmsllinrfzdim, and show that
it is harder than \bmsllinrfqdim, specifically, it is
$\Sigma^P_2$-complete. 
The class $\Sigma^P_2$ is the class of decision problems that can be
solved by a standard, non-deterministic computational model in
polynomial time assuming access to an oracle for an NP-complete
problem. 
I.e., $\Sigma^P_2 = \mbox{NP}^{\mbox{\scriptsize NP}}$.
This class contains both NP and coNP, and is likely to
differ from them both (this is an open problem).


\bthm
\label{thm:dim:int}
For $d\geq 2$, \bmsllinrfzdim is a $\Sigma^P_2$-complete problem.
\ethm 

The rest of this section proves Th.~\ref{thm:dim:int}.
For inclusion in $\Sigma^P_2$ we use a non-deterministic procedure
as in the proof of Th.~\ref{thm:dim:rat}. Note that the procedure needs to find (or
check for existence of) \bmsqlrfs over the integers, so it needs a
coNP oracle.
For $\Sigma^P_2$-hardness we reduce from the canonical
$\Sigma^P_2$-complete problem (follows from
\cite[Th.~4.1]{Stockmeyer77}): evaluation of sentences of the form
\begin{equation} \tag{$\star$}
\abovedisplayskip=4pt
\belowdisplayskip=4pt
\exists X_1 \dots X_n \ \forall X_{n+1} \dots X_{2n} \ \neg \phi(X_1,\dots,X_{2n})
\end{equation}
where the variables $X_i$ are Boolean and the formula $\phi$ is 
in 3CNF form.  Thus, $\phi$ is given as a collection of $m$ clauses,
$C_1,\dots,C_{m}$, each clause $C_i$ consisting of three literals
$L_i^j \in \{ X_1,\dots,X_{2n},\ \neg X_1,\dots, \neg X_{2n} \}$, $1
\le j \le 3$. The reduction is first done for $d=2$, and
later extended to $d>2$ as well.

Let us first explain a well-known approach 
for reducing
satisfiability of a Boolean formula $\phi$ to satisfiability of
integer linear constraints.
We first associate each literal $L_i^j$ with an integer variables
$x_{i,j}$. Note that the same Boolean variable (or its complement)
might be associated with several constraint variables.
Let $C$ be the set of
\begin{inparaenum}[(1)]
\item all conflicting pairs, that is, pairs $((i,j),(r,s))$ such that
  $L_i^j$ is the complement of $L_r^s$; and
\item pairs $((i,j),(i,j'))$ with $1 \le j < j' \le 3$, i.e., pairs of
  literals that appear in the same clause.
\end{inparaenum}
We let $\bintcs$ be a conjunction of the constraints: $x_{i,j}+x_{r,s}\le 1$ for
each $((i,j),(r,s)) \in C$; and $0\le x_{i,j} \le 1$ for each $1\le i
\le m$ and $1\le j\le 3$.
An assignment for $x_{i,j}$ that satisfies $\bintcs$ is called a
\emph{non-conflicting assignment}, since if two variables correspond
to conflicting literals (or to literals of the same clause) they
cannot be assigned $1$ at the same time.
%
%
The next Lemma relates integer assignments with assignments
to the Boolean variables of \EAsentence.  Given a literal $L$, i.e.,
$X_v$ or $\neg X_v$, we let $\litsum(L)$ be the sum of all $x_{i,j}$
where $L_i^j\equiv L$ (we use $0$ and $1$ for \emph{false} and
\emph{true}).

\blem
\label{lem:dim:int:aux}
\textbf{(A)} If $\sigma$ is a satisfying assignment for $\phi$, then
there is a non-conflicting assignment for $\bintcs$ such that
\begin{inparaenum}[(1)]
\item $x_{i,1} + x_{i,2} + x_{i,3}=1$ for all $1 \le i \le m$; 
\item $\sigma(X_v)=1 \Rightarrow \litsum(\neg X_v)=0$; and
\item $\sigma(X_v)=0 \Rightarrow \litsum(X_v)=0$.
\end{inparaenum}
\textbf{(B)} If $\phi$ is unsatisfiable, then for any non-conflicting
assignment for $\bintcs$ there is at least one $1 \le i \le m$ such
that $x_{i,1} + x_{i,2} + x_{i,3}=0$.
\elem

\bprf 
\textbf{(A)} If $\sigma$ satisfies $\phi$, we construct a satisfying
assignment for $\bintcs$: first every $x_{i,j}$ is assigned
the value of $L_i^j$, and then we turn some $x_{i,j}$ from
$1$ to $0$ so that at most one variable of each clause is set to $1$.
Since we only turn $1$s to $0$s, when $\sigma(X_v)=1$
(resp. $\sigma(X_v)=0$) all constraint variables that correspond to
$\neg X_v$ (resp. $X_v$) have value $0$, and thus $\litsum(\neg
X_v)=0$ (resp. $\litsum(X_v)=0$).
\textbf{(B)} If  $\bintcs$ has a non-conflicting assignment in which
$x_{i,1} + x_{i,2} + x_{i,3}=1$ for all $1 \le i \le m$, then we can
construct a satisfying assignment $\sigma$ for $\phi$ in which
$\sigma(X_v)$ is $\max\left( \{ x_{i,j} | L^i_j\equiv X_v \} \cup \{ 1-x_{i,j} |  L^i_j\equiv\neg X_v \} \right)$,
 so $\phi$ is satisfiable.
\eprf

Next we proceed with the reduction, but first we give an outline. 
We build an integer loop, call it $\redmlcloop$, with $2n+2$
abstract transitions: $2n$ transitions named $\chtr{v}{a}$, for $1 \le v \le n$ and $a\in \{0,1\}$;
plus two named $\sattr$ and $\anchortr$.
These are defined so that
existence of a \bmsllrf $\tuple{f_1,f_2}$ for $\redmlcloop$ implies:
\begin{inparaenum}[(1)]
\item $\chtr{v}{0}$ and $\chtr{v}{1}$, for each $1 \le v \le n$,
  cannot be ranked by the same $f_i$, and the order in which they are
  ranked will represent a value for the existentially-quantified
  variable $X_v$;
\item $\sattr$ cannot be ranked by $f_1$, and it is ranked by $f_2$ iff
  $\forall X_{n+1} \dots X_{2n} \ \neg
  \phi(X_1,\dots,X_{2n})$ is true assuming the values induced for
  $X_1,\ldots,X_n$ in the previous step; and
\item $\anchortr$ is necessarily ranked by $f_1$, its only role is to
  force $\sattr$ to be ranked by $f_2$.
\end{inparaenum}
All these points will imply that~\EAsentence is true. For the other
direction, if~\EAsentence is true we show how to construct a \bmsllrf
$\tuple{f_1,f_2}$ for $\redmlcloop$.
Next we formally define the variables and abstract transitions of
$\redmlcloop$, and prove the above claims. 

\medskip
\noindent
\textbf{Variables}: Loop $\redmlcloop$ includes $4m+2n+1$ variables:
\begin{inparaenum}[(1)]
\item every literal $L_i^j$ contributes a variable $x_{i,j}$;
\item for each $1 \le i\le m$, we add a control variable $x_{i,0}$
  which is used to check if clause $C_i$ is satisfied;
\item for each $1 \le v \le n$, we add variables $z_{v,0}$ and
  $z_{v,1}$ which help in implementing the existential quantification;
  and
\item variable $w$, which helps in ranking the auxiliary transition $\anchortr$.
\end{inparaenum}

\medskip
\noindent
\textbf{Transitions}: First we define $\sattr$, the transition that
intuitively checks for satisfiability of $\phi(X_1,\dots,X_{2n})$. It
is a conjunction of the following constraints
{\small
\abovedisplayskip=1pt
\belowdisplayskip=1pt
\begin{align}
  & 0 \le x_{i,j} \le 1\ \land \ x_{i,j}' = x_{i,j}& \mbox{for all } 1\le i \le m ,\ 1\le j\le 3 \label{eq:tr:sat:1}\\[-0.7ex]
  & x_{i,j}+x_{r,s}\le 1 & \mbox{for all }  ((i,j),(r,s)) \in C \label{eq:tr:sat:2}\\[-0.7ex]
  & x_{i,0} \ge 0\ \land\ x'_{i,0} = x_{i,0} + x_{i,1} + x_{i,2} + x_{i,3} - 1 & \mbox{for all } 1\le i \le m \label{eq:tr:sat:3} \\[-0.7ex]
  & z_{v,0} \ge 0\ \land\ z'_{v,0} = z_{v,0} - \litsum(X_v)  & \mbox{for all }  1\le v \le n \label{eq:tr:sat:4} \\[-0.7ex]
  & z_{v,1} \ge 0\ \land\ z'_{v,1} = z_{v,1} - \litsum(\neg X_v)       & \mbox{for all }  1\le v \le n  \label{eq:tr:sat:5}\\[-0.7ex]
  & w' = w  \label{eq:tr:sat:6}
\end{align}
}
Secondly, we define $2n$ transitions which, intuitively, force a
choice of a Boolean value for each of $X_1,\dots,X_n$.  For $1 \le v
\le n$ and $a\in\{0,1\}$, transition $\chtr{v}{a}$ is defined as a
conjunction of the following constraints
{\small
\abovedisplayskip=1pt
\belowdisplayskip=1pt
\begin{align}
& z_{v,a} \ge 0\ \land\ z'_{v,a} = z_{v,a} - 1 \label{eq:tr:ch:1} \\[-0.7ex]
& z_{u,b}  \ge 0  & \mbox{for all } 1\le u \le n, b\in\{0,1\},\ u\ne v \label{eq:tr:ch:2}\\[-0.7ex]
& z'_{u,b} =  z_{u,b}  & \mbox{for all } 1\le u \le n,\ b\in\{0,1\},\ (u,b)\ne (v,a) \label{eq:tr:ch:3} \\[-0.7ex]
& x'_{i,0} \ge 0\ \land\ x'_{i,0} = x_{i,0}\ & \mbox{for all } 1\le i \le m \label{eq:tr:ch:4}\\[-0.7ex]
& w \ge 0\ \land\  w' = w \label{eq:tr:ch:5}
\end{align}
}
%
Finally we define the abstract transition $\anchortr$,
which  aids in forcing a desired form of the \bmsllrf, and it
is defined as a conjunction of the following constraints
{\small
\abovedisplayskip=1pt
\belowdisplayskip=1pt 
\begin{align}
& w \ge 0\ \land\ w' = w - 1 \label{eq:tr:anchor:1}\\[-0.7ex]
& z_{u,b} \ge 0\ \land\ z'_{u,b} =  z_{u,b} & \mbox{for all } 1\le u \le n,\ b\in\{0,1\} \label{eq:tr:anchor:2}
\end{align}
}
Now, we argue that in order to have a two-component \bmsllrf for
$\redmlcloop$, the transitions have to be associated to the two
components in a particular way.

\blem
\label{lem:dim:int:order}
Suppose that $\tuple{f_1,f_2}$ is a \bmsllrf for 
\redmlcloop. Then, necessarily, the correspondence between the
\bmsllrf components and transitions is as follows:
\begin{inparaenum}[(i)]
\item $\anchortr$ is ranked by $f_1$;
\item $\sattr$ is ranked by $f_2$;
\item for $1\le v \le n$, one of $\chtr{v}{0}$ and $\chtr{v}{1}$ is
  ranked by $f_1$, and the other by $f_2$.
\end{inparaenum}
\elem

\bprf 
An \lrf for $\anchortr$ must involve $w$, since it is the only
decreasing variable, and cannot involve any $x_{i,j}$ since
they change randomly. 
%
Similarly, an \lrf for $\sattr$ cannot involve $w$ as it has no lower
bound, and it must involve at least one $x_{i,j}$ since no
function that involves only $z_{v,a}$ variable(s) decreases for an
initial state in which all $x_{i,j}$ are assigned $0$. Note that such
\lrf cannot be non-increasing for $\anchortr$ since $x_{i,j}$ change
randomly in $\anchortr$.
Thus, we conclude that $\anchortr$ must be associated with
$f_1$ and $\sattr$ with $f_2$.
For the last point, for each $1 \le v \le n$, transitions
$\chtr{v}{0}$ and $\chtr{v}{1}$ must correspond to different positions
because variables that descend in one (namely $z_{v,a}$ of $\chtr{v}{a}$) are not bounded in the other 
(since~\eqref{eq:tr:ch:2} requires~$u{\neq}v$). 
%
\eprf

\blem
\label{lem:dim:int:main}
A \bmsllrf of dimension two exists for \redmlcloop iff
\EAsentence is true.
\elem

\bprf
Assume that a \bmsllrf $\tuple{f_1,f_2}$ exists for \redmlcloop, we
show that~\EAsentence is true.
By Lemma~\ref{lem:dim:int:order} we know how the transitions are
associated with the positions, up to the choice of placing
$\Psi_{v,0}$ and $\Psi_{v,1}$, for each $1 \le v \le n$.
Suppose that, for each $1 \le v \le n$, the one which is associated
with $f_2$ is $\Psi_{v,{a_v}}$, i.e., $a_v\in\{0,1\}$, and let
$\cmpl{a}_v$ be the complement of $a_v$.
By construction we know that: 
\begin{inparaenum}[(i)]
\item in $\chtr{v}{a_v}$ the variables $z_{v,{\cmpl{a}_v}}$ and
  $x_{i,j}$ with $j\ge 1$ change randomly, which means that $f_2$
  cannot involve them; and
\item in $\sattr$ the variable $w$ is not lower bounded, which means
  that $f_2$ cannot involve $w$.
\end{inparaenum}
Since these transitions must be ranked by $f_2$, we can assume that
$f_2$ has the form
\(
 f_2(\vec x, \vec z, w) = \sum_i c_{i}\cdot x_{i,0} \, + \, \sum_v c_{v} \cdot z_{v,{a_v}}
\)
where $c_{i}$ and $c_{v}$ are non-negative rational coefficients.
We claim that \EAsentence is necessarily true; for that
purpose we select the value $a_v$ for each $X_v$, and next we show
that this makes it is impossible to satisfy $\phi(X_1,\dots,X_{2n})$.
Assume, to the contrary, that there is a satisfying assignment
$\sigma$ for $\phi$, such that $\sigma(X_v)=a_v$ for all $1\le v \le
n$.
By Lemma~\ref{lem:dim:int:aux} we know that we can construct an
assignment to the variables $x_{i,j}$ such that
\begin{inparaenum}[(i)]
\item $x_{i,1} + x_{i,2} + x_{i,3} = 1$, for each $1 \le i
  \le m$, which means that $x_{i,0}'=x_{i,0}$ at~\eqref{eq:tr:sat:3}; and 
\item for each $1 \le v \le m$, if $a_v=0$ (resp. $a_v=1$), then
  $\litsum(X_v)=0$ (resp. $\litsum(\neg X_v)=0$), which means that
  $z_{v,{a_v}}'=z_{v,{a_v}}$ at~\eqref{eq:tr:sat:4}
  (resp.~\eqref{eq:tr:sat:5}).
\end{inparaenum}
Hence $f_2$ as described above does \emph{not} rank $\sattr$ since
none of its variables change, contradicting our
assumption. We conclude that \EAsentence is true.

Now assume that \EAsentence is true, we construct a
\bmsllrf of dimension two. The assumption means that there are
values $a_1,\dots,a_n$ for the existentially-quantified variables to
satisfy the sentence. Let 
$f_1(\vec x, \vec z, w) = w + \Sigma_{v=1}^n z_{v,{\cmpl{a}_v}}$ and 
$f_2(\vec x, \vec z, w) = \Sigma_{i=1}^m x_{i,0}+\sum_v z_{v,{a_v}}$.
We claim that $\tuple{f_1,f_2}$ is a \bmsllrf such that:
\begin{inparaenum}[(i)]
\item $f_1$ is an \lrf for $\anchortr$ and $\chtr{v}{\cmpl{a}_v}$, and
  non-increasing for $\chtr{v}{a_v}$ and $\sattr$; and 
\item $f_2$ is an \lrf for $\chtr{v}{a_v}$ and $\sattr$.
\end{inparaenum}
All this is easy to verify, except possibly that $f_2$ is an \lrf for
$\sattr$, for which we argue in more detail.  
By assumption, $\phi(a_1,\dots,a_n, X_{n+1},\dots, X_{2n})$ is
unsatisfiable.  Consider a state in which $\sattr$ is enabled; by
(\ref{eq:tr:sat:1},\ref{eq:tr:sat:2}), 
this state may be interpreted as a selection of non-conflicting
literals. 
If one of the selected literals does not agree with the assignment
chosen for $X_1,\dots,X_n$, then by (\ref{eq:tr:sat:4},\ref{eq:tr:sat:5}) the corresponding variable $z_{v,{a_v}}$
is decreasing. Otherwise, there must be an unsatisfied clause, and the
corresponding variable $x_{i,0}$ is decreasing. All other variables
involved in $f_2$ are non-increasing, all are lower bounded, so $f_2$
is an \lrf for $\sattr$.
\eprf

$\Sigma^P_2$-hardness of \bmsllinrfzdim for $d=2$ follows from
Lemma~\ref{lem:dim:int:main}. For $d>2$, we add to \redmlcloop
additional $d-2$ paths as those of Ex.~\ref{ex:bms:maxdim}; and to each
original path in \redmlcloop we add $x'{=}x$ and $y'{=}y$ ($x,y$ are used in
Ex.~\ref{ex:bms:maxdim}). Then, the new loop has a \bmsllrf of
dimension $d$ iff~\EAsentence is true. This concludes the proof of
Th.~\ref{thm:dim:int}.
%
%
%


\section{Related Work}
\label{sec:rw}

\llrfs appear in the classic works of Turing~\cite{Turing48} and
Floyd~\cite{Floyd67}.
Automatic generation of \lrfs and \llrfs
for linear-constraint loops begins, in the context of logic programs,
with Sohn and van Gelder~\cite{DBLP:conf/pods/SohnG91}.
For imperative programs, it begins with Col{\'o}n and Sipma
\cite{DBLP:conf/tacas/ColonS01,DBLP:conf/cav/ColonS02}.
The work of Feautrier on scheduling~\cite{Feautrier92.1,Feautrier92.2}
includes, in essence, generation of \lrfs and \llrfs.
All these works gave algorithms that yield polynomial time complexity
(inherited from LP), except for Col{\'o}n and Sipma's method which is
based on LP duality and polars. The polynomial-time LP method later
reappeared
in~\cite{DBLP:journals/tplp/MesnardS08,DBLP:conf/vmcai/PodelskiR04}.
These methods are complete over the rationals and can be used in an
integer setting by relaxing the loop from integer to rational
variables, sacrificing completeness.
This completeness problem was pointed out (but not solved) in
\cite{DBLP:journals/tplp/MesnardS08,Rybalchenko04},
while~\cite{CookKRW10,Feautrier92.1} pointed out the role of the
integer hull in ensuring completeness. Bradley et
al.~\cite{DBLP:conf/concur/BradleyMS05} use a bisection search over
the space of coefficients for inferring \lrfs over the integers, which yields
completeness at exponential cost (as argued in~\cite{Ben-AmramG13jv}).

Alias et al.~\cite{ADFG:2010} extended the LP approach to \llrfs,
obtaining a polynomial-time algorithm which is sound and complete over
the rationals (for their notion of \llrf).  
The (earlier) work of Bradley et al.~\cite{DBLP:conf/cav/BradleyMS05}
introduced \bmsllrfs and used a ``constraint-solving method" that
finds such \llrfs along with supporting invariants. The method
involves an exponential search for the association of paths to \llrf
components, and is complete over the reals.
Subsequent work used more complex extensions of the \llrf
concept~\cite{DBLP:conf/icalp/BradleyMS05,DBLP:conf/vmcai/BradleyMS05}.
Harris et al.~\cite{harris2011alternation} demonstrate that it is
advantageous, to a tool that is based on a CEGAR loop, to search for
\llrfs instead of \lrfs only. The \llrfs they use are \bmsllrfs.
Similar observations have been reported in
\cite{DBLP:conf/tacas/CookSZ13} (also using \bmsllrfs),
\cite{DBLP:conf/cav/BrockschmidtCF13} (using \adfgllrfs) and
\cite{LarrazORR13} (using a an iterative construction that extends
\bmsllrfs).
Heizmann and Leike~\cite{HeizmannLeike:TACAS2014} generalize the
constraint-based approach by defining the concept of a ``template" for
which one can solve using a constraint solver. They also provide a
template for \adfgllrfs (of constant dimension).
Ben-Amram~\cite{DBLP:journals/corr/abs-1105-6317} shows that
\emph{every} terminating monotonicity-constraint program has a
\emph{piecewise} \llrf of dimension at most $2n$. Piecewise \llrfs are
also used in~\cite{DBLP:conf/esop/UrbanM14}, with no completeness
result, there they are inferred by abstract interpretation.


\section{Conclusion}
\label{sec:conclusion}

This work contributes to understanding the design space of the
ranking-function method, a well-known method for termination analysis
of numeric loops, as well as related analyses (iteration bounds,
parallelization schedules).  This design space is inhabited by several
kinds of ``ranking functions'' previously proposed.
We focused on \bmsllrfs and compared them to other proposals of a
similar nature. We characterized the complexity of finding, or
deciding the existence of, \bmsllrf for rational and integer \mlc
loops.
We also compared these three methods regarding the dimension of the
\llrf, and the complexity of optimizing the dimension, which turns out
to be essentially harder for \bmsllrfs.
Given our reductions, it is easy to show that it is impossible to
approximate the minimal dimension of \bmsllrfs, in polynomial time,
within a factor \emph{smaller than} $\frac{3}{2}$, unless
$P{=}\mathit{NP}$ for rational loops, and $\Sigma^P_2 {=} \Delta^P_2$ for
integer loops (see~\apprefapproxalg).

We conclude that none of the three methods is universally preferable.
Even \adfgllrfs, which in principle are weaker than both other methods,
have an advantage, in that the algorithm for computing them may be more efficient
in practice (due to solving smaller LP problems). If this is not a concern,
they can be replaced by \bgllrfs, so we are left with two, incomparable techniques.
This incomparability stems from the fact that \bgllrfs and \bmsllrfs
relax the restrictions of \adfgllrfs in two orthogonal directions: the
first in quantifying over concrete transitions rather than abstract
ones, and the second in allowing negative components.
By making both relaxations, we get a new type of
\llrf~\cite{LarrazORR13}, which is as in Def.~\ref{def:bgllrf} but
relaxing condition~\eqref{eq:bg:llrf2} to hold only for $j=i$, 
but for which the computational complexity questions are still open.

\bibliographystyle{plain}
\bibliography{integer-loops}

\ifTR
\appendix
\newpage


\newcommand{\bmsqlrfp}[1]{\ensuremath{\bmsqlrf({#1})}}

\section{\bmsllinrfz is coNP-complete}
\label{sec:BMS-coNP}

The coNP-hardness follows from the reduction
in~\cite[Sect.~3.1]{Ben-AmramG13jv}, since it constructs a loop that
either does not terminate or has an \lrf.
Next we prove inclusion in coNP by showing that the complement
problem, i.e., the nonexistence of a \bmsllrf, has a polynomially
checkable witness.
We assume a given \mlc loop $\transitions_1,\ldots,\transitions_k$
where each $\transitions_i$ is given as a set of linear constraints,
over $2n$ variables ($n$ variables and $n$ primed variables).
In this appendix we assume familiarity with Section 2.1
of~\cite{Ben-AmramG13jv} (preliminaries on polyhedra).

Recall that Proposition~\ref{prop:rat:llrfcond}, when applied to
$\intpoly{\transitions_1},\ldots,\intpoly{\transitions_k}$, implies
that $\intpoly{\transitions_1},\ldots,\intpoly{\transitions_k}$ has no
\bmsllrf iff there is a subset of the transition polyhedra that has no
\bmsqlrf.
This suggests that this subset can be used as a witness for the
nonexistence of a \bmsllrf. However, checking that such a subset has
no \bmsqlrf cannot be done in polynomial time using the Algorithm of
Lemma~\ref{lem:int:bmsqlrfalg}, since it requires computing the
corresponding integer hull, and thus cannot be directly used as a
witness.
Instead, we show that there is finite set of integers points, related
to this subset of the transition polyhedra, that can witness the
nonexistence of a \bmsqlrf, and, moreover, can be checked in
polynomial time (by checking that some corresponding set of
constraints has no solution, over the rationals).
Without loss of generality, assume that the subset of the transition
polyhedra that we are considering, for the nonexistence of \bmsqlrf,
is $\intpoly{\transitions_1},\ldots,\intpoly{\transitions_\ell}$ for
some $\ell\le k$.

We first show that there is a polynomially checkable witness for the
nonexistence of a \bmsqlrf for
$\intpoly{\transitions_1},\ldots,\intpoly{\transitions_\ell}$ that
ranks a specific $\intpoly{\transitions_p}$ for $1\le p \le
\ell$ --- we refer to such \bmsqlrf as $\bmsqlrfp{p}$. Then we use
this witness to construct one for the non-existence of \bmsqlrf.

\bdfn
\label{def:bms:noqlrf:witness}
Let $X=X_1\cup\cdots\cup X_\ell$, $Y=Y_1\cup\cdots\cup Y_\ell$ and $1
\le p \le \ell$, such that
\begin{inparaenum}[\upshape(\itshape a\upshape)]
\item\label{def:bms:noqlrf:witness:1} $X_i \subseteq \intpoly{\transitions_i}$; 
\item\label{def:bms:noqlrf:witness:2} $Y_i\subseteq \intpoly{\recess{\transitions_i}}$;
\item\label{def:bms:noqlrf:witness:3} $Y_i\neq\emptyset \Rightarrow X_i\neq\emptyset$; and
\item\label{def:bms:noqlrf:witness:4} $X_p \neq \emptyset$.
\end{inparaenum}
We say that $\tuple{X,Y}$ is a witness against the existence of a
\bmsqlrfp{p} for
$\intpoly{\transitions_1},\ldots,\intpoly{\transitions_\ell}$ if the
following set of linear constraints has no solution
\begin{subequations}
\begin{align}
\vect{\rfcoeff}\cdont \vec{x} + {\rfcoeff}_0 \ge 0  
   &~~~ \mbox{ for all } \vec{x}'' \in X_p   \label{eq:bms:noqlrfp:1}\\
\vect{\rfcoeff}\cdot(\vec{x} - \vec{x}')  \ge 1  
&~~~ \mbox{ for all } \vec{x}'' \in X_p \label{eq:bms:noqlrfp:2}\\
\vect{\rfcoeff}\cdot(\vec{x} - \vec{x}')  \ge 0
&~~~ \mbox{ for all } \vec{x}'' \in X_i ~~~(i\ne p) \label{eq:bms:noqlrfp:3}\\
\vect{\rfcoeff}\cdont\vec{y} \ge 0 
&~~~ \mbox{ for all } \vec{y}'' \in Y_p \label{eq:bms:noqlrfp:4}\\
\vect{\rfcoeff}\cdot (\vec{y} - \vec{y}')  \ge 0 
&~~~ \mbox{ for all } \vec{y}'' \in Y_i  ~~~ \forall i \label{eq:bms:noqlrfp:5}
\end{align}
\end{subequations}
The variables in the above constraints are
${\rfcoeff}_0,\vect{\rfcoeff}$, and they are rational-valued.
\edfn

\blem 
\label{lem:bms:noqlrf:witness:1}
Let $X=X_1\cup\cdots\cup X_\ell$, $Y=Y_1\cup\cdots\cup Y_\ell$ and
$1\le p\le \ell$ be as in
Definition~\ref{def:bms:noqlrf:witness}. Then
$\intpoly{\transitions_1},\cdots,\intpoly{\transitions_\ell}$
has no \bmsqlrfp{p}.
\elem

\bprf
Assume the contrary, i.e., there is $({\rfcoeff}_0,\vect{\rfcoeff})
\in \rats^{n+1}$ such that
$\rho(\vec{x})=\vect{\rfcoeff}\cdot\vec{x}+{\rfcoeff}_0$ is a
\bmsqlrfp{p} for
$\intpoly{\transitions_1},\cdots,\intpoly{\transitions_\ell}$.
By assumption, \eqref{eq:bms:noqlrfp:1}-\eqref{eq:bms:noqlrfp:5} has
no solution, hence, they are not satisfied by the specific
$({\rfcoeff}_0,\vect{\rfcoeff})$ that we have chosen above.
But \eqref{eq:bms:noqlrfp:1}-\eqref{eq:bms:noqlrfp:3} are clearly
satisfied because $\rho$ is a \bmsqlrfp{p}, and thus one of
\eqref{eq:bms:noqlrfp:4} or \eqref{eq:bms:noqlrfp:5} is not
satisfied. We reason on these two cases separately.

\medskip
\noindent
\emph{Case}~1: Suppose \eqref{eq:bms:noqlrfp:5} is not satisfied, for
some $\vec{y}''\in Y_p$. That is,
$\vect{\rfcoeff} \cdot \vec{y} < 0$.
Choose $\vec{x}''\in X_p$, and note that for any integer $a\ge 0$, the
integer point $\vec{z}''=\vec{x}'' + a\cdot\vec{y}''$ is a transition
in $\intpoly{\transitions_p}$, and
$\vec{z}''=\tr{\vec{x}\phantom{'}+a\cdot\vec{y}}{\vec{x}'+a\cdot\vec{y}'}$.
Now,
\[
\rho(\vec{z}) = \vect{\rfcoeff} \cdot  (\vec{x} + a\cdot\vec{y}) + \rfcoeff_0 =
\rho(\vec{x}) + a\cdot(\vect{\rfcoeff} \cdot  \vec{y})
\]
It is easy to see that for sufficiently large $a$ we get
$\rho(\vec{z})<0$, since $\vect{\rfcoeff} \cdot \vec{y} < 0$, which
contradicts that $\rho$ is \bmsqlrfp{p}.

\medskip
\noindent
\emph{Case}~2: Suppose \eqref{eq:bms:noqlrfp:4} is not satisfied, for
some $\vec{y''}\in Y_i$. That is,
$\vect{\rfcoeff} \cdot (\vec{y} - \vec{y}') < 0$.
Choose $\vec{x}''\in X_i$ and define $\vec{z}''$ as above. Now,
\[
\rho(\vec{z})-\rho(\vec{z}') 
=
   \vect{\rfcoeff} \cdot  ((\vec{x} + a\cdot\vec{y}) - (\vec{x}' + a\cdot\vec{y}')) 
=   \rho(\vec{x})-\rho(\vec{x}') + a\cdot(\vect{\rfcoeff} \cdot  (\vec{y}-\vec{y}')) \\
\]
It is easy to see that for sufficiently large integer $a$ we get
$\rho(\vec{z})-\rho(\vec{z}') < 0$, since $\vect{\rfcoeff} \cdot
(\vec{y} - \vec{y}') < 0$, which contradicts that $\rho$ is
\bmsqlrfp{p}.
This concludes the proof.
\eprf

\blem 
\label{lem:bms:noqlrf:witness:2}
If there no \bmsqlrfp{p} for
$\intpoly{\transitions_1},\cdots,\intpoly{\transitions_\ell}$, then
there are finite sets $X=X_1\cup\cdots\cup X_\ell$ and
$Y=Y_1\cup\cdots\cup Y_\ell$, 
 fulfilling the
conditions of Definition~\ref{def:bms:noqlrf:witness}.
\elem

\bprf 
For $1\le i\le \ell$, let $\inthull{\transitions_i} = \convhull\{X_i\}
+ \cone\{Y_i\}$ be the generator representation of the integer hull of
$\transitions_i$, and define $X=X_1\cup\cdots\cup X_\ell$ and
$Y=Y_1\cup\cdots\cup Y_\ell$.
We claim that $\tuple{X, Y}$, 
fulfill the
conditions of Definition~\ref{def:bms:noqlrf:witness}. Assume the
contrary, i.e., \eqref{eq:bms:noqlrfp:1}-\eqref{eq:bms:noqlrfp:5} has
a solution $({\rfcoeff}_0,\vect{\rfcoeff}) \in \rats^{n+1}$, we show
that $\rho(\vec{x})=\vect{\rfcoeff}\cdot\vec{x}+{\rfcoeff}_0$ is a
\bmsqlrfp{p}, contradicting the assumption that no \bmsqlrfp{p}
exists.

Pick a point $\vec{x}''\in\intpoly{\transitions_i}$, and let
$X_i=\{\vec{x}''_1,\ldots,\vec{x}''_m\}$ and
$Y_i=\{\vec{y}''_1,\ldots,\vec{y}''_t\}$.
Note that $\vec{x}''=\sum_{i=1}^m a_i \cdot \vec{x}''_i + \sum_{j=1}^t
b_j \cdot \vec{y}''_j$ for some rationals $a_i,b_j\ge 0$, where
$\sum_{i=1}^m a_i = 1$. We show that $\rho$ correctly ranks
$\vec{x}''$, i.e., fulfills the corresponding conditions of \bmsqlrf
depending on if $\vec{x}''$ comes from $\intpoly{\transitions_p}$ or
from $\intpoly{\transitions_i}$ with $i\neq p$:

\begin{itemize}
\item If $\vec{x}''\in \intpoly{\transitions_p}$, then each $x''_j\in
  X_i$ satisfies (\ref{eq:bms:noqlrfp:1},\ref{eq:bms:noqlrfp:2}) and
  each $y''_j\in Y_i$ satisfies
  (\ref{eq:bms:noqlrfp:4},\ref{eq:bms:noqlrfp:5}), then, it is easy to
  check that this necessarily imply $\rho(\vec{x})\ge 0$ and
  $\rho(\vec{x})-\rho(\vec{x}')\ge 1$.
\item If $\vec{x}''\not\in \inthull{\transitions_p}$, then each
  $x''_j\in X_i$ satisfies \eqref{eq:bms:noqlrfp:3} and each $y''_j\in
  Y_i$ satisfies \eqref{eq:bms:noqlrfp:5}, it is easy to check that
  this necessarily imply $\rho(\vec{x})-\rho(\vec{x}')\ge 0$.
\end{itemize}
This concludes the proof.
\eprf

\blem
\label{lem:nollrf:witness-size}
If there is a finite witness for the nonexistence of \bmsqlrfp{p} for
$\intpoly{\transitions_1},\ldots,\intpoly{\transitions_\ell}$, then
there is one whose bit-size is polynomial in the bit-size of
${\transitions_1},\ldots,{\transitions_\ell}$.
\elem

\bprf
By Lemma~\ref{lem:bms:noqlrf:witness:2}, we conclude that if there is
a witness then there is one, $X=X_1\cup\cdots\cup X_\ell$ and
$Y=Y_1\cup\cdots\cup Y_\ell$, such that $X_i$ and $Y_i$ come from the
generator representation of $\inthull{\transitions_i}$.

Recall that \eqref{eq:bms:noqlrfp:1}-\eqref{eq:bms:noqlrfp:5} has no
solution for the points of $X$ and $Y$.
A corollary of Farkas' Lemma~\cite[p.~94]{Schrijver86} states that if
a finite set of inequalities over $\rats^d$, for some $d>0$, has no
solution, there is a subset of at most $d+1$ inequalities that has no
solution.
Since the set of inequalities
\eqref{eq:bms:noqlrfp:1}-\eqref{eq:bms:noqlrfp:5} is over
$\rats^{n+1}$, there is a subset of at most $n+2$ inequalities that
has no solution.

These inequalities correspond to $n+2$ points out of the sets $X_i$,
$Y_i$. Let $\hat X_i$ (respectively $\hat Y_i$) be the set of points
that come from $X_i$ (respectively $Y_i$). 
Since \eqref{eq:bms:noqlrfp:1}-\eqref{eq:bms:noqlrfp:5} has no
solution for these sets, at least one of the points must come from a
set $\hat X_p$ (otherwise $\vec{0}$ is a solution).  But
$n+1$ other points might come from sets $\hat Y_i$. Since a witness
must satisfy $\hat Y_i\ne \emptyset \Rightarrow \hat X_i\ne \emptyset$
and $\hat X_p\neq 0$, we may have to add $n+1$ points to form a valid
witness, for a total of $2n+3$.
The bit-size of this witness is polynomial in the input bit-size,
because each point comes from the generator representation of some
$\inthull{\transitions_i}$, and it is known that it is possible to
choose a generator representation in which each vertex has a bit-size
that is polynomial in the bit-size of $\transitions_i$
(see~\cite[Th.~2.7 and Th.~2.8]{Ben-AmramG13jv}).
\eprf

Checking that a given $X=X_1\cup\cdots\cup X_\ell$ and
$Y=Y_1\cup\cdots\cup Y_\ell$ is a witness as in
Definition~\ref{def:bms:noqlrf:witness} can be done in polynomial time
as follows: First we verify that each $\vec{x}''\in X_i$ is in
$\intpoly{\transitions_i}$, which can be done by verifying $A_i
\vec{x}'' \le \vec{c}_i$; and that each $\vec{y}''\in Y_i$ is in
$\intpoly{\recess{\transitions_i}}$, which can be done by verifying
$A_i \vec{y} \le \vec{0}$. This is done in polynomial time.
Note that according to Lemma~\ref{lem:bms:noqlrf:witness:1} it is not
necessary to check that $X_i$ and $Y_i$ come from a particular
generator representation. Then we check that
\eqref{eq:bms:noqlrfp:1}-\eqref{eq:bms:noqlrfp:5} has no solution,
which can be done in polynomial time since it is an LP problem over
$\rats^{n+1}$.

\bcor
\label{cor:bms:noqlrf:witness}
There is a polynomially checkable witness for the nonexistence of a
\bmsqlrf for
$\intpoly{\transitions_1},\ldots,\intpoly{\transitions_\ell}$.
\ecor

\bprf 
The witness consists of $\ell$ witnesses,
$\tuple{X^1,Y^1},\ldots,\tuple{X^\ell,Y^\ell}$, each as in
Definition~\ref{def:bms:noqlrf:witness} for some $1 \le p\le \ell$.
Thus, the $i$-th one witnesses against the existence of
$\bmsqlrfp{i}$. Thus all together witness against the existence of
\bmsqlrf.
Its size is clearly polynomial in the the input-bit size, and it can
be checked in polynomial time by checking each $\tuple{X^i,Y^i}$ as
described before.
\eprf

\bthm
\label{th:llrf:conp-inc}
$\bmsllinrfz \in \mathrm{coNP}$ for \mlc loops.
\ethm

\bprf 
Straightforward, given Proposition~\ref{prop:rat:llrfcond} and
Corollary~\ref{cor:bms:noqlrf:witness}.
\eprf


\newcommand{\nocommonqlrfwit}[4]{\Gamma(#1,#2,#3,#4)}
\newcommand{\nolrfwit}[2]{\Psi(#1,#2)}

\section{Complexity of the bounded-dimension decision problem for \adfgllrf and \bgllrf}
\label{sec:adfg-bg-complexity}

We assume a given \mlc loop $\transitions_1,\ldots,\transitions_k$
where each $\transitions_i$ is given as a set of linear constraints
over $2n$ variables ($n$ variables and $n$ primed variables).
The different bounded-dimension
decision problems are denoted, naturally, by \bgllinrfqdim, \bgllinrfzdim,
\adfgllinrfqdim, and \adfgllinrfzdim. In this appendix we assume familiarity with sections 2.1 and 5 of~\cite{Ben-AmramG13jv}.

\bthm 
\label{thm:length:adfg-bg:rat}
\bgllinrfqdim and \adfgllinrfqdim are in P.
\ethm

\bprf 
We solve the problem by synthesizing an optimal-dimension \bgllrf or
\adfgllrf, which in both cases is PTIME. Then, we simply answer
positively if and only if we found a tuple of dimension at most $d$.
\eprf

Next we move to \bgllinrfzdim and \adfgllinrfzdim, and show that both
are coNP-complete.
In both cases coNP-hardness is straightforward, since for $d=1$ it
becomes the problem of deciding if there is an \lrf, and the argument can easily be
extended to larger $d$. The rest of this
section is dedicated to the inclusion in coNP.

\bthm 
\label{thm:length:adfg-bg:int}
\bgllinrfzdim and \adfgllinrfzdim are in coNP.
\ethm

\noindent
We prove for \bgllinrfzdim, and then
comment on how the proof can be adapted to \adfgllinrfzdim as well.

The main step of the proof is to describe the form of a witness
\emph{against} the existence of a $d$-component \bgllrf.  The
technical details of the proofs can be worked out exactly as in the
corresponding proofs in Appendix~\ref{sec:BMS-coNP} of this article,
or in~\cite[Sec.~5.2]{Ben-AmramG13jv}. Thus,
we only sketch them here.

\blem 
\label{lem:nollrf:witness-1}
Let 
\[
 T_d \subseteq T_{d-1} \subseteq \dots \subseteq T_1 \subseteq \intpoly{\transitions_1}\cup\cdots\cup\intpoly{\transitions_k} ,
\]
such that 
\begin{inparaenum}[\upshape(\itshape i\upshape)]
\item\label{def:nollrf:sets:1} there is no \lrf for $T_d$; and
\item\label{def:nollrf:sets:2} for each $\ell = 1,\dots,d-1$,
  \textbf{every quasi-\lrf} for $T_\ell$ does not decrease on any of the
  transitions $T_{\ell + 1}$.
\end{inparaenum}
Then $\intpoly{\transitions_1},\ldots,\intpoly{\transitions_k}$
has no \bgllrf of dimension (at most) $d$.
Conversely, if there is no \bgllrf of dimension at most $d$, there is a
chain of sets as above.  \elem

\bprf 
$(\Rightarrow)$ Suppose in contradiction that
$\llrfsym=\tuple{\rho_1,\ldots,\rho_d}$ is a \bgllrf (note that we can
always pad the tuple to dimension $d$ if it is of a smaller dimension).
Then $\rho_1$ is a quasi-\lrf for $T_1$, and so by
(\ref{def:nollrf:sets:2}) does not decrease on $T_2$. Hence
$\tuple{\rho_2,\ldots,\rho_d}$ is a \bgllrf for $T_2$.  Proceedings in
this way we deduce that $\rho_d$ must be an \lrf for $T_d$,
contradicting (\ref{def:nollrf:sets:1}).

$(\Leftarrow)$ Suppose that there is no \bgllrf of dimension at most $d$.
Following the \bgllrf (synthesis) algorithm~\cite[Alg.~1,
p.30]{Ben-AmramG13jv}, we see that one of the following must happen:
(1) within $d$ recursive calls, the algorithm fails to find a
non-trivial quasi-\lrf, or (2) a $d+1$ recursive call is reached.
We construct sets $T_1,\ldots,T_d$ that
satisfy~(\ref{def:nollrf:sets:1},\ref{def:nollrf:sets:2}) as follows:
Let $\tuple{\poly{P}_{j1},\ldots,\poly{P}_{jk}}$, for $1 \le j \le d$,
be the parameters received by the \bgllrf algorithm in $j$-th
invocation (if the algorithm stops at iteration $s < d$, we assume
$\poly{P}_{ji}=\poly{P}_{si}$ for any $s < j \le d$); and define
$T_j=\cup_{i=1}^k \intpoly{\poly{P}_{ji}}$, for $1 \le j \le d$.
\eprf

In what follows, given sets of integer points $X' \subseteq X$ and $Y'
\subseteq Y$, we let $\nocommonqlrfwit{X}{Y}{X'}{Y'}$ be the
conjunction of the following inequalities:
\begin{subequations}
\begin{align}
\vect{\rfcoeff}\cdont \vec{x} + {\rfcoeff}_0 \ge 0  
   &~~~ \mbox{ for all } \vec{x}'' \in X   \label{eq:noQRF:1}\\
\vect{\rfcoeff}\cdont\vec{y} \ge 0 
&~~~ \mbox{ for all } \vec{y}'' \in Y \label{eq:noQRF:2}\\
\vect{\rfcoeff}\cdot(\vec{x} - \vec{x}')  \ge 0  
&~~~ \mbox{ for all } \vec{x}'' \in X \label{eq:noQRF:3}\\
\vect{\rfcoeff}\cdot (\vec{y} - \vec{y}')  \ge 0 
&~~~ \mbox{ for all } \vec{y}'' \in Y \label{eq:noQRF:4}\\
\sum_{\vec{x}'' \in X'} \vect{\rfcoeff}\cdot (\vec{x} - \vec{x}') \,+&\sum_{\vec{y}'' \in Y'} \vect{\rfcoeff}\cdot(\vec{y} - \vec{y}')  \ge 1  \label{eq:noQRF:5}
\end{align}
\end{subequations}
Intuitively, $\tuple{X,Y}$ and $\tuple{X',Y'}$ will be generators of sets
of integer points $T' \subseteq T$ such that the solutions of
$\nocommonqlrfwit{X}{Y}{X'}{Y'}$ are the quasi-\lrfs of $T$ that also
decrease on some points of $T'$.
For sets of integer points $X$ and $Y$ we let $\nolrfwit{X}{Y}$ be the
conjunction of the following inequalities:
\begin{subequations}
\begin{align}
\vect{\rfcoeff}\cdont \vec{x} + {\rfcoeff}_0 \ge 0  
   &~~~ \mbox{ for all } \vec{x}'' \in X   \label{eq:noLRF:1}\\
\vect{\rfcoeff}\cdont\vec{y} \ge 0 
&~~~ \mbox{ for all } \vec{y}'' \in Y \label{eq:noLRF:2}\\
\vect{\rfcoeff}\cdot(\vec{x} - \vec{x}')  \ge 1  
&~~~ \mbox{ for all } \vec{x}'' \in X \label{eq:noLRF:3}\\
\vect{\rfcoeff}\cdot (\vec{y} - \vec{y}')  \ge 0 
&~~~ \mbox{ for all } \vec{y}'' \in Y \label{eq:noLRF:4}
\end{align}
\end{subequations}
Intuitively, $\tuple{X,Y}$ will generate a set of integer
points $T$, and the solutions of $\nolrfwit{X}{Y}$ are all \lrfs of
$T$.

\bdfn
\label{def:nollrf:witness}
Given $\tuple{X_1,Y_1},\ldots,\tuple{X_{d},Y_{d}}$, where $X_j=X_{j
  1}\cup\cdots\cup X_{j k}$ and $Y_j=Y_{j 1}\cup\cdots\cup Y_{j k}$,
such that
\begin{enumerate}[\upshape(\itshape a\upshape)]
\item\label{def:nollrf:witness:1} $X_{d i} \subseteq X_{(d-1) i} \subseteq \dots \subseteq X_{1 i} \subseteq \intpoly{\transitions_i}$;
\item\label{def:nollrf:witness:2} $Y_{d i} \subseteq Y_{(d-1) i} \subseteq \dots \subseteq Y_{1 i} \subseteq \intpoly{\recess{\transitions_i}}$; 
\item\label{def:nollrf:witness:3} $Y_{j i}\neq\emptyset \Rightarrow X_{j i}\neq\emptyset$.
\end{enumerate}
We say that $\tuple{X_1,Y_1},\ldots,\tuple{X_{d},Y_{d}}$ form a
witness against the existence of a \bgllrf of dimension at most $d$ for
$\intpoly{\transitions_1},\ldots,\intpoly{\transitions_k}$ if it
satisfies the following requirements:
\begin{enumerate}[\upshape(\itshape a\upshape)]
\setcounter{enumi}{3}
\item\label{def:nollrf:witness:4} $\nolrfwit{X_{d}}{Y_{d}}$ has no solution; and
\item\label{def:nollrf:witness:5}
  $\nocommonqlrfwit{X_i}{Y_i}{X_{i+1}}{Y_{i+1}}$, for any $1 \le i \le d-1$, has no solution.
\end{enumerate}
\edfn

Each $\tuple{X_j,Y_j}$ corresponds to a set of
integer points $T_j$ such that $T_{j+1} \subseteq T_{j}$. In addition,
condition~\eqref{def:nollrf:witness:4} guarantees that $T_d$ has no
\lrf, and condition~\eqref{def:nollrf:witness:5} guarantees that there
is no quasi-\lrf for $T_j$ that is decreasing for some points of
$T_{j+1}$.

\blem 
\label{lem:nollrf:witness-2}
Let $\tuple{X_1,Y_1},\ldots,\tuple{X_{d},Y_{d}}$ be as in
Definition~\ref{def:nollrf:witness}. Then there are $T_d \subseteq
\dots \subseteq T_1\subseteq
\intpoly{\transitions_1}\cup\cdots\cup\intpoly{\transitions_k}$
satisfying the requirements of Lemma~\ref{lem:nollrf:witness-1}.
\elem

\bprf
We construct sets of transitions 
$T_d \subseteq T_{d-1} \subseteq \cdots \subseteq T_{1} \subseteq
\intpoly{\transitions_1}\cup\cdots\cup\intpoly{\transitions_k}$
that satisfy the requirements of Lemma~\ref{lem:nollrf:witness-1}.
We construct $T_j$ from $\tuple{X_j,Y_j}$ as follows: 
\[
T_j = \{ \vec{x}''+a\vec{y}'' \mid \vec{x}''\in X_{ji}, \vec{y}''\in
Y_{ji}, \mbox { integer $a\ge 0$ } \}\; .
\] 
Note that for $\vec{x}''\in X_{ji}$ and $\vec{y}''\in Y_{ji}$, the
point $\vec{x}''+a\vec{y}''$, for any integer $a\ge 0$, is a
transition in $\intpoly{\transitions_i}$, thus $T_j \subseteq
\intpoly{\transitions_1}\cup\cdots\cup\intpoly{\transitions_k}$.
We claim that these sets satisfy the requirements of
Lemma~\ref{lem:nollrf:witness-1}; the proof can be worked out
similarly to~\cite[Lemma 5.18]{Ben-AmramG13jv}).
\eprf

The last result states that our witnesses are sound---they really
imply that there is no \bgllrf of the desired dimension.  Next we should
also prove that when there is no such \bgllrf, witness sets as above
exist, and their size can be polynomially bounded.

\blem 
\label{lem:llrf:witness:2}
Suppose that
$\intpoly{\transitions_1},\ldots,\intpoly{\transitions_k}$ has no
\bgllrf of dimension at most $d$.  Then there are
$\tuple{X_1,Y_1},\ldots,\tuple{X_{d},Y_{d}}$ of bit-size polynomially
bounded by the bit-size of the input transition polyhedra (as
constraints), fulfilling the conditions of
Definition~\ref{def:nollrf:witness}.
\elem

\bprf 
Consider again the \bgllrf
algorithm~\cite[Alg.~1,p.30]{Ben-AmramG13jv}, if the integer loop
$\intpoly{\transitions_1},\ldots,\intpoly{\transitions_k}$ has no
\bgllrf of dimension at most $d$, one of the following happens:
(1) within $d$ recursive calls, the algorithm fails to find a
non-trivial quasi-\lrf, or (2) a $d+1$ recursive call is reached.
Let $\tuple{\poly{P}_{j1},\ldots,\poly{P}_{jk}}$, for $1 \le j \le d$,
be the parameters received by the \bgllrfs algorithm in $j$-th
recursive call (if the algorithm stops at iteration $s < d$, we let
$\poly{P}_{ji}=\poly{P}_{si}$ for any $s < j \le d$).
Define $T_j=\cup_{i=1}^k \intpoly{\poly{P}_{ji}}$, for all $1 \le j
\le d$. Then, clearly $T_1,\ldots,T_d$ are sets of transitions that
satisfy the requirements of Lemma~\ref{lem:nollrf:witness-1}.  We
construct a witness that corresponds to these sets as follows:
First note that each $\poly{P}_{ji}$ is integral, and has a
corresponding generator representation
$$\poly{P}_{ji}= \convhull\{X_{ji}\} + \cone\{Y_{ji}\} \,.$$
where $X_{ji}$ and $Y_{ji}$ are finite sets of integer points.  
Then, we define each component $\tuple{X_j,Y_j}$ of the witness as
$X_j=X_{j1}\cup\ldots\cup X_{jk}$ and $Y_j=Y_{j1}\cup\ldots\cup
Y_{jk}$.

This witness satisfies condition~\eqref{def:nollrf:witness:3} of
Definition~\ref{def:nollrf:witness}, because we may assume that none
of the transition polyhedra is a cone (otherwise the loop clearly does not terminate),
 and thus $X_{ij}\neq\emptyset$.
To show that it satisfies
conditions~(\ref{def:nollrf:witness:1},\ref{def:nollrf:witness:2}) as
well, we rely on the following fact~\cite[p.107]{Schrijver86}: if a
polyhedron $\poly{P}=\convhull\{X\}+\cone\{Y\}$ is a face of a
polyhedron $\poly{P}'=\convhull\{X'\}+\cone\{Y'\}$, then $X\subseteq
X'$ and $Y\subseteq Y'$.
Now a property of the \bgllrf algorithm~\cite[Lemma
5.8]{Ben-AmramG13jv} is that $\poly{P}_{(j+1) i}$ is a  
face of $\poly{P}_{j i}$, and thus $X_{(j+1)i}
\subseteq X_{j i}$ and $Y_{(j+1)i} \subseteq Y_{j i}$, so the witness
satisfies
conditions~(\ref{def:nollrf:witness:1},\ref{def:nollrf:witness:2}).
Showing that
conditions~(\ref{def:nollrf:witness:4},\ref{def:nollrf:witness:5})
hold can be worked out as in~\cite[Lemma~5.19]{Ben-AmramG13jv}.
Finally, We can reduce the witness above to polynomial bit-size, using
the same arguments as~\cite[Lemma~5.22]{Ben-AmramG13jv}.
\eprf

Checking a witness can be done in polynomial time as follows: First we
verify that each $\vec{x}''\in X_{ji}$ is in
$\intpoly{\transitions_i}$, which can be done by verifying $A_i
\vec{x}'' \le \vec{c}_i$; and that each $\vec{y}''\in Y_{ji}$ is in
$\intpoly{\recess{\transitions_i}}$, which can be done by verifying
$A_i \vec{y}'' \le \vec{0}$. This is done in polynomial time.
Checking that $\nolrfwit{X_{d}}{Y_{d}}$ and
$\nocommonqlrfwit{X_i}{Y_i}{X_{i+1}}{Y_{i+1}}$ have no solution can be
done in polynomial time since it is an LP problem over the
rationals. This concludes the proof of
Theorem~\ref{thm:length:adfg-bg:int}, and thus \bgllinrfzdim is
coNP-complete.

\subsection{The case of \adfgllrfs}

Next we explain how to adapt the above proof to \adfgllrfs. The same approach
works for adapting the coNP-completeness proof
of~\cite{Ben-AmramG13jv} for the existence of \bgllrfs to the case of
\adfgllrfs.

The important difference between the quasi-\lrfs used in \adfgllrf
from those of \bgllrfs is that they the must be non-negative over all
$\intpoly{\transitions_1},\ldots,\intpoly{\transitions_k}$. This means
that when checking that a witness has no solution (signifying that there is
no \lrf, or no quasi-\lrf with a certain non-triviality restriction)
we should take this additional restriction into
account. This can be done by extending the witness with an extra
component $X'=X_1'\cup\cdots\cup X_k'$ and $Y'=Y_1'\cup\cdots\cup
Y_k'$, such that
\begin{inparaenum}[\upshape(\itshape i\upshape)]
\item $X'_{i} \subseteq \intpoly{\transitions_i}$ and finite; 
\item $Y'_{i} \subseteq \intpoly{\recess{\transitions_i}}$ and finite;
  and
\item $Y'_{i}\neq\emptyset \Rightarrow X_{i}'\neq\emptyset$.
\end{inparaenum}
In addition, we add the the following in inequalities requirements from a witness
\begin{subequations}
\begin{align}
\vect{\rfcoeff}\cdont \vec{x} + {\rfcoeff}_0 \ge 0  
   &~~~ \mbox{ for all } \vec{x}'' \in X' \\
\vect{\rfcoeff}\cdont\vec{y} \ge 0 
&~~~ \mbox{ for all } \vec{y}'' \in Y' \,.
\end{align}
\end{subequations}
%


\section{Approximation of Minimum Dimension}
\label{sec:approx-alg}

Since finding out whether a \bmsllrf of dimension $d$ exists is
NP-hard and $\Sigma^P_2$-hard, for rational and integer loops,
respectively, a natural question to ask is if we can approximate the
minimum dimension in polynomial time.

\bigskip
\noindent
\emph{Rational loops.}
It is know that it is impossible to approximate, in polynomial time,
the chromatic number of $r$-uniform hypergraphs on $n$ vertices within
a factor $n^{1-\eps}$, for any $\eps > 0$, unless $\mathit{NP}
\subseteq \mathit{ZPP}$~\cite{KrivelevichS2003}.  Given our reduction,
we conclude that it is impossible to approximate the minimal dimension
of \bmsllrfs within a factor $k^{1-\eps}$, for any $\eps > 0$, unless
$\mathit{NP} \subseteq \mathit{ZPP}$ (recall that $n$ vertices
generate $k$ paths in our reduction).
Similarly, we cannot do such approximation within a factor
\emph{smaller than} $\frac{3}{2}$, unless $P=\mathit{NP}$, because we
can then decide if a $3$-uniform hypergraph has $2$-coloring, which is
an NP-complete~\cite{Lovasz:hypergraphs1973}.

\paragraph{Integer loops.}
A polynomial algorithm, even given access to an NP oracle (a SAT
solver) for free, cannot approximate the minimum dimension $d$ of
\bmsllrfs within a factor \emph{smaller than} $\frac{3}{2}$, unless
$\Sigma^P_2 = \Delta^P_2$. This is because if such an algorithm
exists, then for the loop $\redmlcloop$, a result of $2$ will mean
that~\EAsentence is true, and any other result (necessarily 3 or 4,
since it has to be under $\frac{3}{2} \cdot 3$) will mean that it is
false.  Thus a $\Sigma^P_2$-hard problem is solved in $\Delta^P_2$
complexity.

\fi

\end{document}